\documentclass[aps,pre,longbibliography,superscriptaddress,10pt]{revtex4-1}

\usepackage{graphicx}
\usepackage{color}
\usepackage{subfigure}
\usepackage{amsmath}
\usepackage{amssymb} 
\usepackage{color}

\usepackage[margin=1.25in]{geometry}

\begin{document}

\title{Growth of clogs in parallel microchannels}

\author{Alban Sauret}
\affiliation{Department of Mechanical Engineering, University of California, Santa Barbara, California, USA}
\affiliation{Surface du Verre et Interfaces, UMR 125, CNRS/Saint-Gobain, Aubervilliers, France}

\author{Katarzyna Somszor}
\affiliation{Department of Mechanical and Aerospace Engineering, NYU Tandon School of Engineering, Brooklyn, New York, USA}

\author{Emmanuel Villermaux}
\affiliation{Aix-Marseille Universit\'{e}, CNRS, Centrale Marseille, IRPHE, Marseille, France}
\affiliation{Institut Universitaire de France, Paris, France}
\affiliation{CNRS/MIT/AMU Joint Laboratory MultiScale Materials Science for Energy and Environment, MIT Energy Initiative, Massachusetts Institute of Technology, Cambridge, Massachusetts, USA}

\author{Emilie Dressaire} \email{emilie.dressaire@u-psud.fr}
\affiliation{Department of Mechanical Engineering, University of California, Santa Barbara, CA, USA}
\affiliation{Department of Mechanical and Aerospace Engineering, NYU Tandon School of Engineering, Brooklyn NY, USA}
\affiliation{FAST, UMR 7608, Universit\'e Paris-Sud, CNRS, Universit\'{e} Paris-Saclay, Orsay, France}

\date{\today}

\begin{abstract}
During the transport of colloidal suspensions in microchannels, the deposition of particles can lead to the formation of clogs, typically at constrictions. Once a clog is formed in a microchannel, advected particles form an aggregate upstream from the site of the blockage. This aggregate grows over time, which leads to a dramatic reduction of the flow rate. In this paper, we present a model that predicts the growth of the aggregate formed upon clogging of a microchannel. We develop an analytical description that captures the time evolution of the volume of the aggregate, as confirmed by experiments performed using a pressure-driven suspension flow in a microfluidic device. We show that the growth of the aggregate increases the hydraulic resistance in the channel and leads to a drop in the flow rate of the suspensions. We then derive a model for the growth of aggregates in multiple parallel microchannels where the clogging events are described using a stochastic approach. The aggregate growths in the different channels are coupled. Our work illustrates the critical influence of clogging events on the evolution of the flow rate in microchannels. The coupled dynamics of the aggregates described here for parallel channels is key to bridge clogging at the pore scale with macroscopic observations of the flow rate evolution
at the filter scale.
\end{abstract}

\maketitle

\section{Introduction}

The transport of suspensions in microchannels and pores is a standard process in biological systems and industrial and environmental applications such as water filtration \cite{yao1971water} and fracking \cite{Marder_13,bocquet_16}. The transport can be associated with the contamination of tubings or soils through particle deposition \cite{blazejewski1997soil,reddi2000permeability}. Clogging occurs when one or several deposited particles prevent the advection of other particles beyond the clogging site. The fomation of a clog suddenly decreases the flow rate of the suspension. Hence, the clogging dynamics is important to a wide range of applications of suspension flow in porous media, from fractured rocks to filters and sensors and in biophysical systems \cite{blazejewski1997soil,reddi2000permeability,huang2014highly,pang2015deformability,guo2017microfluidic,bacher2017clustering}. Many studies have investigated the trapping of particles leading to the clog formation. Early studies characterized particle trapping and clogging in filters through flow rate measurements. The efficiency and lifetime of a filter was obtain though theoretical models and/or macroscopic measurements \cite{ruth1933studies,ruth1935studies,coulson1968flow,goldsztein2004suspension,duclos2006three,dalwadi2016multiscale,sanaei2017flow,han2017network}. Different theoretical models have been developed considering various blocking mechanisms and some predictions of the time evolution of the flow rate were provided \cite{Hermia}. More recently, microfluidic tools have allowed the description of clog formation at the pore scale \cite{dressaire2017clogging}. In a microchannel, stable clogs often forms at constrictions, where the width of the channel locally decreases. Microfluidic systems allow us to record directly the clogging events as well as the time evolution of the filtered particles.

\smallskip

The clogging of a microchannel follows three possible mechanisms depending on the size of the particles, their surface chemistry or charge and the concentration of the suspension \cite{dressaire2017clogging}. For dense suspensions of particles without surface charges, bridging, i.e., the formation of an arch of particles is the main clogging mechanism \cite{ramachandran1999plugging,sharp2005flow,zuriguel2014clogging,pacheco,endo2017obstacle,gella2017linking,marin2018clogging,hidalgo2018flow}. This situation is similar to what is observed at a silo outlet with dry granular material \cite{zuriguel2014invited}. Bridging of particles is unlikely to occur with dilute suspensions of particles and sufficiently wide constriction \cite{marin2018clogging}. However, particles smaller than the constriction, with charged surfaces, are still able to clog the device by successive deposition on the walls \cite{lin2009particle,bacchin2011colloidal,agbangla2014collective,bacchin2014clogging,massenburg2016clogging}. Experimentally, Wyss \textit{et al.} \cite{wyss2006mechanism} have observed that particle-wall interactions can determine the clogging dynamics of parallel microchannels. Recent studies have investigated the deposition of particles prior to clogging \cite{dersoir2015clogging,de2016dynamics,dersoir2017clogging,cejas2017particle}. Finally, a last possible mechanism is sieving: A single large particle blocks the channel through size exclusion \cite{sauret2014clogging,yoon2016clogging}. Different strategies are now available to avoid, control, and/or leverage clogging by rigid or deformable particles such as dust particles, colloids, cells, or emulsions drops. In addition, sieving-based methods are also used for biological screening \cite{huang2014highly,pang2015deformability,guo2017microfluidic}. Although the clogging mechanisms can differ, the following evolution of the aggregates will be the same. Here, we rely on clogging by sieving that allows a well-defined theoretical description of the clogging probability. In addition, the sieving by large particles is a common process when using polydisperse suspensions.

\smallskip

After the clog formation, additional particles are advected toward the clog where they stop and form an aggregate that grows over time. The aggregate decreases the flow rate in the device over time \cite{brenner1961three}. Also, the presence of an aggregate can redistribute the flow in the neighboring channels \cite{liot2017pore}. This coupling leads to a complex clogging and flow dynamics in porous media \cite{van2018cooperative}. Such interplay is particularly important to understand how the results obtained at the pore-scale level can be applied to predict the behavior of parallelized or macroscopic systems. This is relevant to filters and to the fracking industry when trying to estimate the flow rate and production volume \cite{Marder_13}. The modification of the flow features could also add uncertainty to the screening of biological cells.

\smallskip

In this paper, we aim to describe the clogging dynamics and the time evolution of the flow rate in parallel microchannels. To achieve this description, we first consider experimentally and theoretically the time evolution of the aggregate and then use this description in a numerical model. We rely on experiments performed with a suspension of monodisperse particles containing a minute amount of contaminants whose size is about 5 to 10 times the diameters of the particles in suspension. The contaminants are responsible for the clogging of microchannels by sieving, which is described by a Poisson distribution \cite{sauret2014clogging}. In the first part of the paper, using clear microfluidic devices, with a single channel, we measure the time evolution of the aggregate for a pressure-driven flow and model the evolution of both the flow rate and the particle aggregate. The experimental results are successfully compared to the analytical model and used to obtain the permeability of the aggregate. In the second part of our work, we extend our model to the clogging and aging of more complex and realistic systems with multiple parallel channels clogged by sieving. Such systems are used for the filtration of polluted water or biological samples through membranes. The analytical prediction of the mean clogging time and the mean lifetime of the device are compared to numerical results. The model also allows us to describe the time evolution of the flow rate in the entire device.

\section{Experimental observations}

\subsection{Experimental methods}

\begin{figure}
  \begin{center}
\subfigure[]{ \includegraphics[width=8cm]{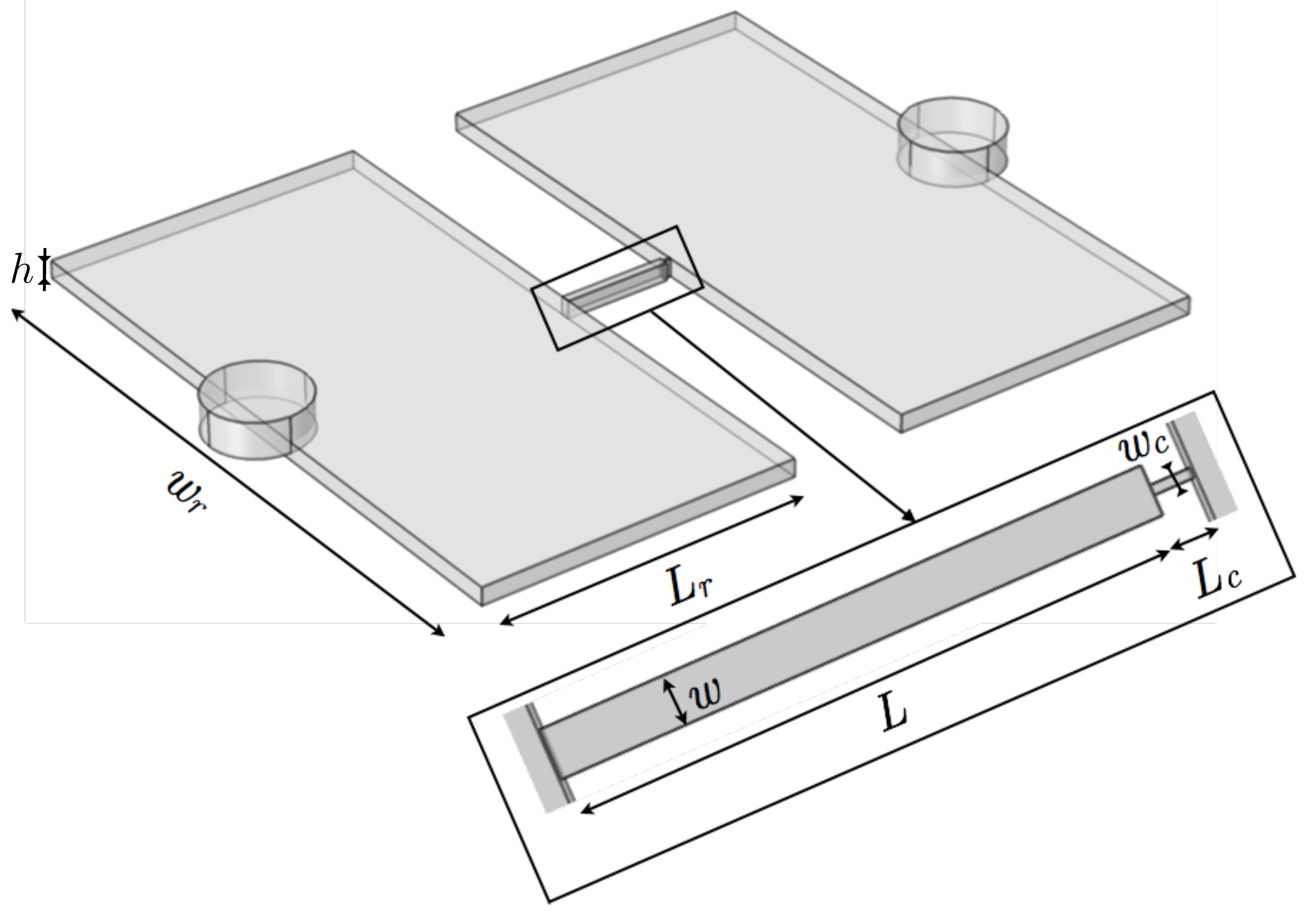}} \qquad
 \subfigure[]{\includegraphics[width=5cm]{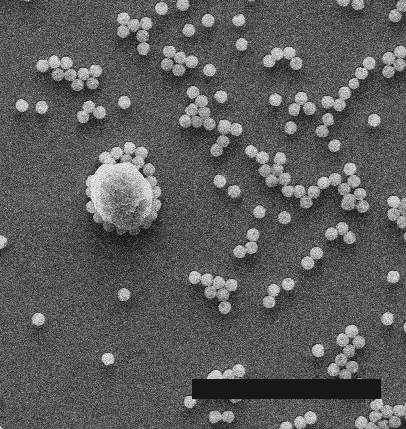}}
    \caption{(a) Schematic of the microfluidic device with one channel. (b) SEM images of the colloidal particles used in the experiments showing the presence of a large contaminant ($d_c \simeq 12\,\mu{\rm m}$). Scale bar is $30\,\mu{\rm m}$.}
    \label{setup}
  \end{center}
\end{figure}

The microfluidic devices used in this study are fabricated using soft lithography methods. The four walls of a device are made of polydimethylsiloxane (PDMS) to ensure that the surface properties are identical. The microfluidic system has a constant height $h=14.2\,\mu{\rm m}$ and is shown in Fig. \ref{setup}(a). The inlet and outlet reservoirs of width $w_{r}=3.6\,{\rm mm}$ and length $L_{r}=1.8\,{\rm mm}$, are connected to a pressurized syringe containing the suspension and to the atmosphere, respectively. The two reservoirs are connected by either a single microchannel [see Fig. \ref{setup}(a)] or 40 parallel identical microchannels [see Fig. \ref{exp_40canaux}]. Each microchannel consists of a channel of length $L$ and a small constriction where a contaminant gets trapped, forming a clog. The channel, of width $w=40$ or $60\,\mu{\rm m}$ and length $L=600\,\mu{\rm m}$, gets filled with the particles that form the aggregate after the clogging occurs at the entrance of the constriction. The constriction, or narrow section, has a width $w_{c}=10\,\mu{\rm m}$ and a length $L_c=40\,\mu{\rm m}$.

A difference of pressure $\Delta p \in [6.9,\,48.2]\,{\rm kPa}$ drives the flow through the device. The flow rate $Q$ is proportional to the pressure drop through the total hydraulic resistance of the device $R_{tot}$: $\Delta p=R_{tot}\,Q$. We flow an aqueous suspension of polystyrene particles (Polysciences Inc.) of mean diameter $2.1\,\mu{\rm m}$. Within the time scale of an experiment, typically 1 h, we do not observe any noticeable adhesion between particles or between a particle and a PDMS wall 
\cite{sauret2014clogging}.
Therefore, the particles mainly interact through steric effects. The suspension has a small polydispersity (coefficient of variance $5\%$) and contains a tiny amount of large contaminants which are not perfectly spherical but have one dimension $d_c$ larger than the width of the constriction $w_{c}$, as illustrated by the scanning-electron-microscope (SEM) image in Fig. \ref{setup}(b). The concentration of large contaminants is equal to $c =(5.6 \pm 1.1) \times 10^8 \,{\rm m^{-3}}$ for a $2\times 10^{-3}$ v/v suspension \cite{sauret2014clogging}. Therefore, the fraction of large contaminants in the suspension is $f_c=c/c_{particles}=1.4 \pm 0.3 \times 10^{-6}$. We perform experiments varying the volume fraction of particles in the suspension $\alpha$ in the range $[0.1\%,\,0.4\%]$.

 We use digital video microscopy to record movies of the microfluidic channel and measure the dynamics of the aggregate at the single-pore level \cite{wyss2006mechanism,dersoir2015clogging}. The typical transit time of a particle through the entire microfluidic device is around $0.1$ s. Thus, at standard video rates, we do not image individual particles passing through the channel, only the growth of the aggregate. 

\subsection{Phenomenology}

\begin{figure}[h!]
  \begin{center}
\includegraphics[width=14cm]{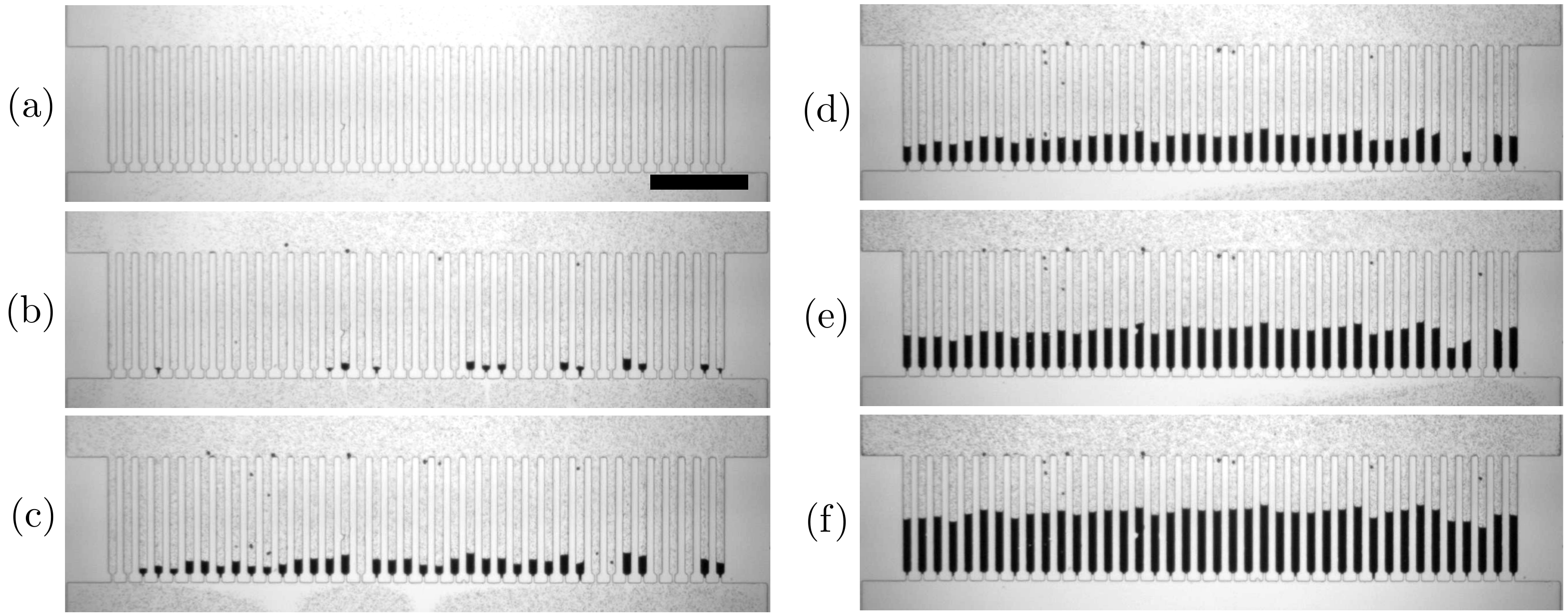}
    \caption{(a) Time evolution of clogging by sieving and the growth of the aggregates in 40 parallel microchannels taken at (a) $t=5 \,{\rm s}$, (b) $t=25 \,{\rm s}$, (c) $t=60 \,{\rm s}$, (d) $t=120 \,{\rm s}$, (e) $t=200 \,{\rm s}$, and (f) $t=400 \,{\rm s}$. The flow of the colloidal suspension is from top to bottom and the experimental parameters are $\Delta p=13.8\,{\rm kPa}$, $\alpha=0.4\,\%$, $w=40\,\mu{\rm m}$, and $N=40$. The scale bar is 500 $\mu{\rm m}$.}
    \label{exp_40canaux}
  \end{center}
\end{figure}

As an illustration of the situation we aim to describe in this paper, we first consider the microfluidic device in which 40 parallel and identical microchannels connect the two reservoirs. This situation is often used to investigate clogging in microfluidic devices: A difference of pressure $\Delta p$ drives the suspension into the inlet reservoir and then through $N$ parallel channels ($N=40$, here). This method is generally used to collect a significant amount of data in a relatively short time at the pore scale so that reliable statistics can be obtained. An example of a time series of such experiment is shown in Figure \ref{exp_40canaux}.

We observe a relatively fast clogging of the first channels, and then the following channels clog one after the other, at a slower pace. The pores that are first clogged are randomly distributed [Figs. \ref{exp_40canaux}(b) and \ref{exp_40canaux}(c)]. This observation is consistent with a stochastic process: A given number of colloidal particles need to flow through each channel before a large contaminant arrives at a constriction \cite{sauret2014clogging}. The last channels are longer to clog, as seen in Figs. \ref{exp_40canaux}(d) and \ref{exp_40canaux}(e), and as previously reported for other clogging mechanisms \cite{wyss2006mechanism}. After a channel clogs, we observe the growth of an aggregate because of the accumulation of colloidal particles that can no longer flow through the constriction. The lengths of all aggregates seem to become relatively uniform over time [Fig. \ref{exp_40canaux}(f)]. When some microchannel(s) are still open, the aggregates already formed start growing. Such observation indicates that the usual assumption of a zero-flow rate in a clogged microchannel cannot describe the evolution of the system accurately. The aggregate in a clogged channel keeps growing showing that the flow rate remains non zero in a clogged channel. To obtain a more realistic model, we need to consider the influence of the permeability of the aggregate on the flow rate and the clogging dynamics.

\subsection{Experimental growth of an aggregate in a single micro-channel}

\begin{figure}[h!]
  \begin{center}
\includegraphics[width=14cm]{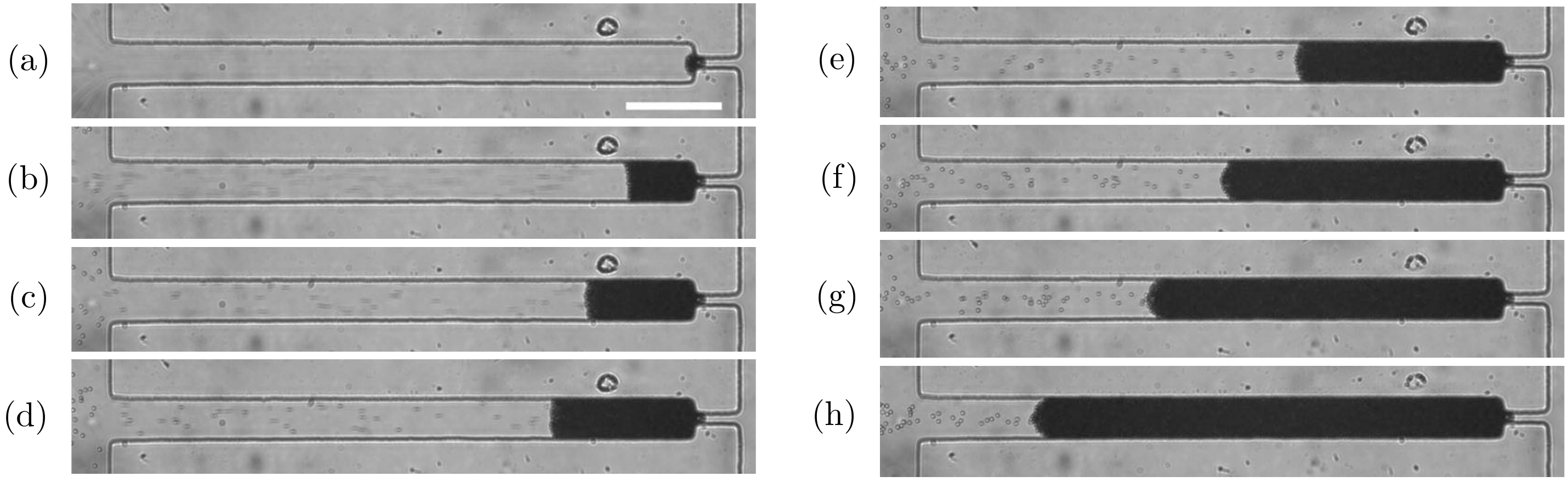}
    \caption{Time evolution of the aggregate following the initial clogging in a single microchannel device with $\Delta p=13790$ Pa and $\alpha=0.2\,\%$. The contaminants clogged the constricting channel at (a) $t=0$s. The later particles build an aggregate: (b) $t=50$s, (c) $t=150$s, (d) $t=350$s, (e) $t=950$s, (f) $t=1950$s, (g) $t=2950$s and (h) $t=4450$s. Scale bar is $100\,\mu$m.}
    \label{exp_1}
  \end{center}
\end{figure}

To model the growth of the aggregates in $N$ parallel channels and the evolution of the total flow rate we first need to understand the growth dynamics of an aggregate in a single channel [Fig. \ref{setup}(a)]. We will then use the model obtained for a single channel to describe the evolution of $N$ microchannels in parallel. We experimentally consider the clogging and the resulting growth of an aggregate in a channel of width $w$, as shown in Fig. \ref{exp_1}. We observe that the growth of the aggregate is fast during the first seconds, then slows down over time. This dynamics is induced by an increase in the hydraulic resistance that accompanies the growth of the colloidal aggregate. From these experiments, we can extract the volume of the colloidal aggregate in the channel,  $V_{agg}$. In the following, we will use the length of the aggregate $\ell(t) = V_ {agg}(t) / (w \, h)$ to describe the growth dynamics.

\begin{figure*}
  \begin{center}
 \subfigure[]{\includegraphics[width=0.32\textwidth]{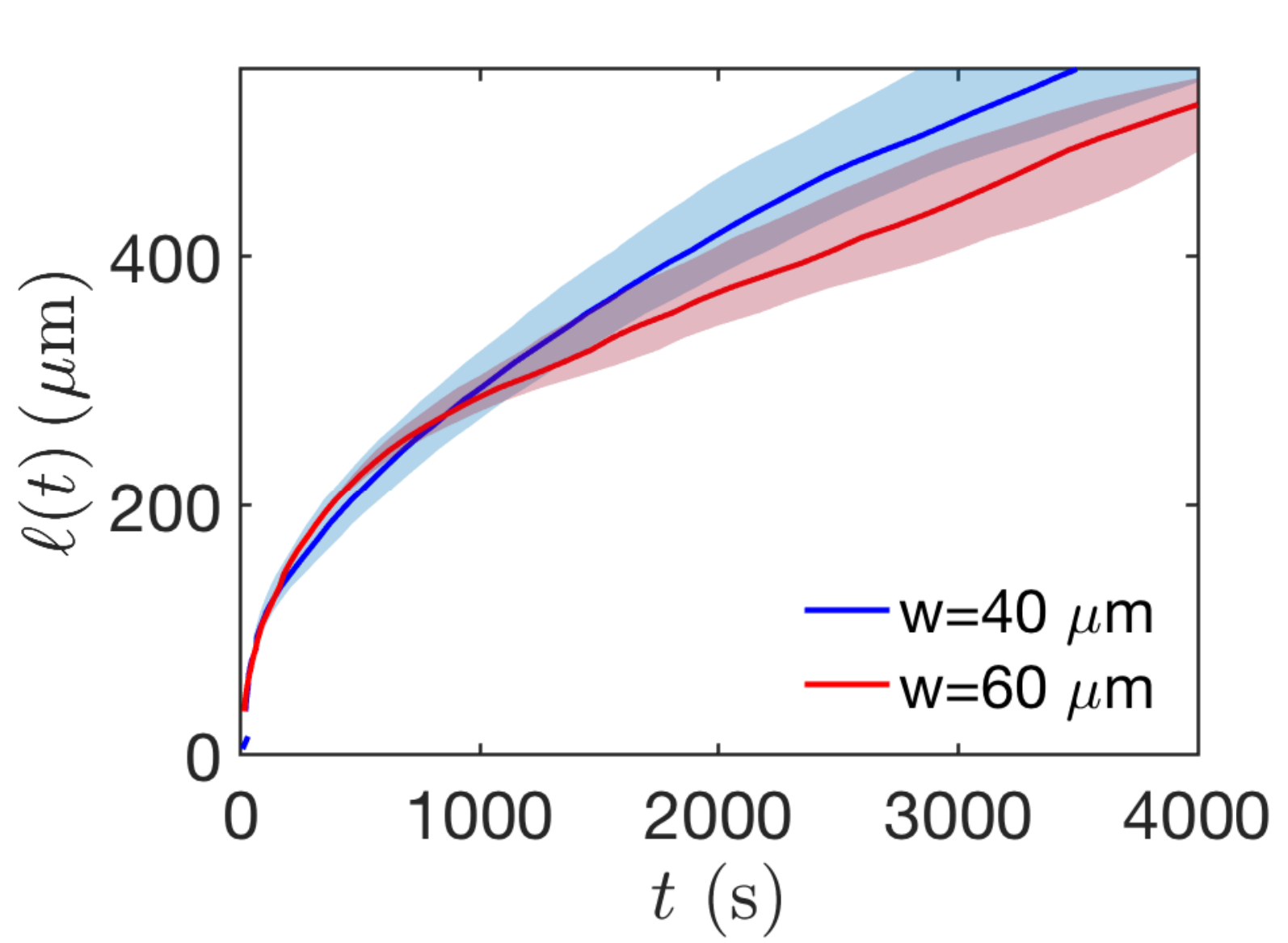}}
 \subfigure[]{\includegraphics[width=0.32\textwidth]{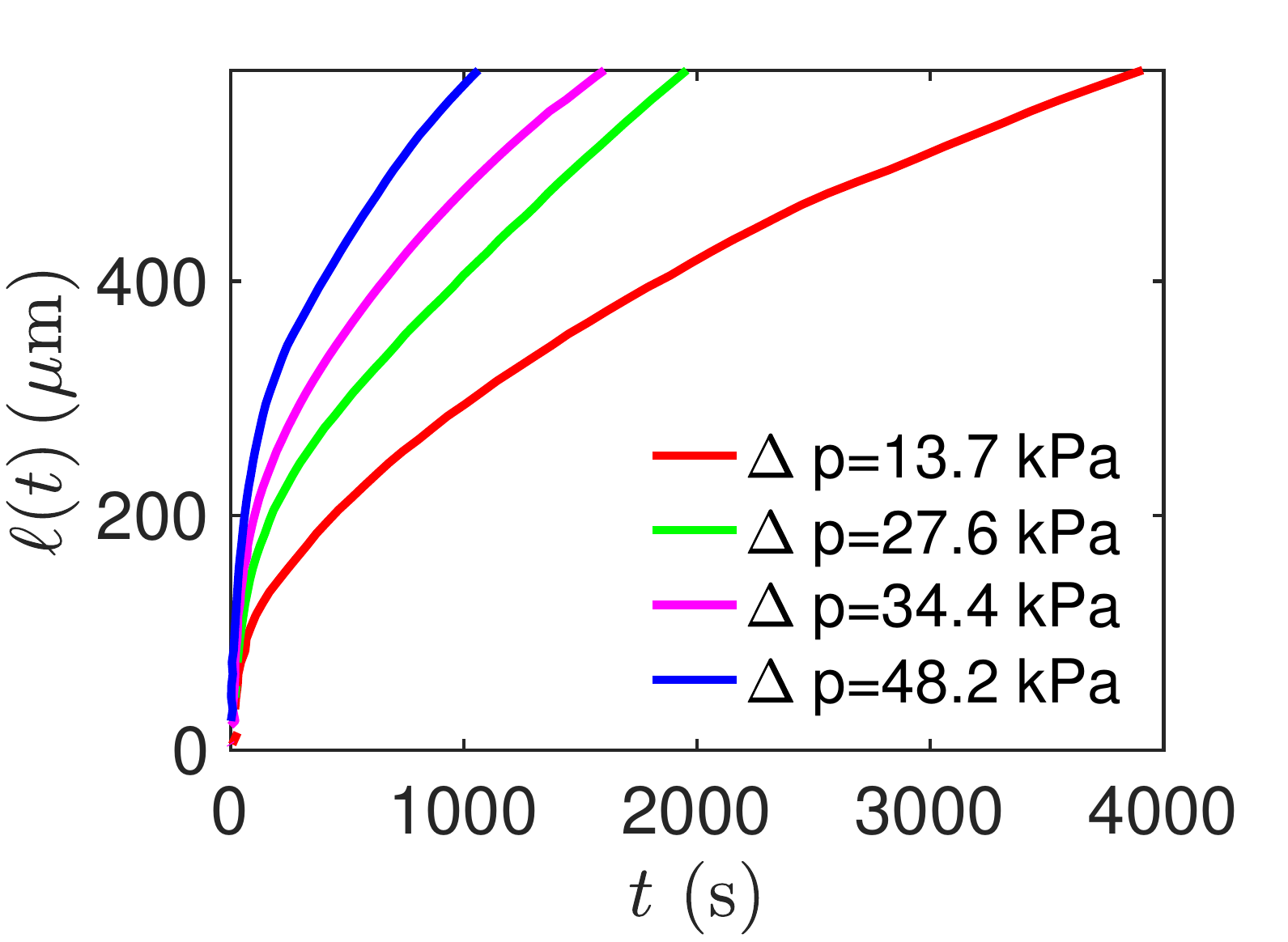}}
 \subfigure[]{\includegraphics[width=0.32\textwidth]{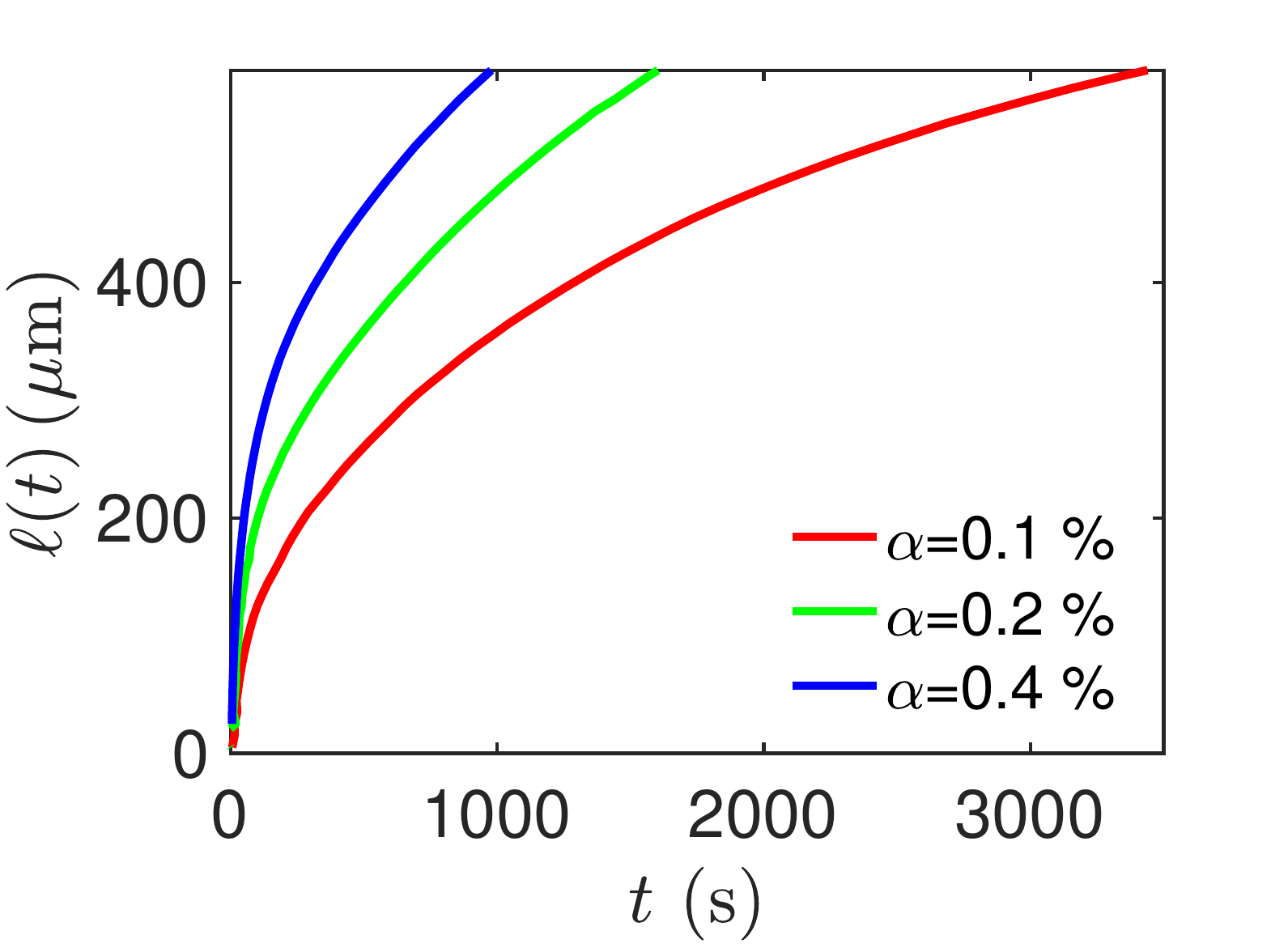}}
    \caption{Time evolution of the length of the aggregate for (a) two different widths $w$, $\Delta p=13.7\,{\rm kPa}$ and $\alpha=0.2\,\%$, the colored regions show the standard deviation over all the averaged experiments, (b) different pressure $\Delta p$, a width of the channel $w=40\,\mu{\rm m}$ and a concentration $\alpha=0.2\,\%$, and (c) different concentrations $\alpha$, $w=40\,\mu{\rm m}$ and $\Delta p=34.4\,{\rm kPa}$.}
    \label{syst_exp}
  \end{center}
\end{figure*}

We systematically study the growth of the aggregate in the single channel. Each configuration is run at least five times to average the growth dynamics. Small variations can be observed between two experiments because of the different positions of the contaminant that blocks the constriction.  The time evolution of the length of the aggregate is plotted in Figs. \ref{syst_exp}(a) and \ref{syst_exp}(c). The growth of the aggregate is initially fast before slowing down when the length $\ell(t)$ increases. We characterize the influence of the different parameters on the aggregate growth.

We first investigate the influence of the width of the microchannel $w$ [Fig. \ref{syst_exp}(a)]. The aggregate length has a similar growth dynamics for the two different widths considered here, $w = 40 \, \mu {\rm m} $ and $ w = 60 \, \mu {\rm m} $, but the volume of the aggregate $ V_{agg} = \ell (t) \, w \, h $ increases faster for a wider channel. When varying the pressure difference $\Delta p$, we observe that as $\Delta p$ increases, the growth of the aggregate is faster [Fig. \ref{syst_exp}(b)]. The flow rate in a single microchannel $q$ is related to the pressure difference $\Delta p$ via the expression $\Delta p = R_{tot} \, Q $. Thus, for a given aggregate length, i.e., a total hydraulic resistance $R_{tot}[\ell (t)] $, a larger value of $\Delta p$ will result in a larger flow rate $q$. Therefore, more particles contribute to the growth of the aggregate at any time step. A more concentrated suspension leads to a faster clogging but also to a faster aggregate growth [Fig. \ref{syst_exp}(c)]. The flow rate $q$ at a given aggregate length $\ell(t) $ is independent of the volume fraction of particles $\alpha$ and only depends on the pressure difference, which remains constant. However, the amount of particles that contribute to the aggregate, $ V_{p} = \alpha \, V $ ($V_p$ is the volume of particles and $ V $ is the volume of fluid), increases with $\alpha$.

In the next section, we develop a model to describe the growth of the aggregate for a single channel. This model captures the dependence of the growth on the parameters of the system ($\Delta p$, $\alpha$ and $w$). Once benchmarked, we shall use this model to describe the evolution of the filtered flow rate in a microfluidic device composed of $ N $ microchannels in parallel.


\section{Growth of the aggregate in a single microchannel} 

\subsection{Evolution of the length of the aggregate}

We consider a device composed of two reservoirs and a single microchannel with a constriction, as illustrated in Fig. \ref{setup}(a). The flow rate through the microchannel $q_1$ is equal to the flow rate in the entire device, $q_1=Q$ 
\cite{sauret2014clogging}. The average time needed for one contaminant to clog the microchannel is $\langle t_{clog,1}\rangle=1/(q_1\,c)$, where $c$ is the number concentration of contaminants. After a contaminant gets trapped at time $t_0$, the particles accumulate in the channel and an aggregate build up, which increases the resistance of the channel and decreases the flow rate.

To model the growth of the aggregate, we consider the total volume of particles that has flowed through the channel after clogging:
\begin{equation} 
V(t)=\int_{t_0}^t Q\,\rm{d}t.
\end{equation}
The corresponding volume of the aggregate is ${V}_{agg}(t)=h\,w\,\ell(t)={V_{p}(t)}/{\phi}$, where $V_{p}(t)=\alpha\,V(t)$ is the volume of particles, $\phi$ is the compacity (or packing fraction) of the aggregate and $\alpha$ is the volume
fraction of particles in the suspension. The volume of the aggregate and time $t$ is thus ${V}_{agg}(t)=\alpha\,V(t)/\phi$, which leads to the length of the aggregate:
\begin{equation}  \label{evolution_l}
\ell(t)=\frac{\alpha\,\Delta p}{\phi\,h\,w}\int_{t_0}^t \frac{\rm{d}t}{R_{tot}}.
\end{equation} 
The Reynolds number associated to the fluid flow in the aggregate is $Re_p=u\,d/\nu$, where $u$ is the mean flow velocity, $d$ is the diameter of the particles and $\nu$ is the kinematic viscosity of the suspension. Here, $Re_p \ll 1$ and we use Darcy's law that describes the flow through a porous media of permeability $k$:
\begin{equation} \label{Darcy}
\frac{\Delta p'}{\ell(t)}=-\frac{\mu}{k}\,\frac{{Q}}{h\,w},
\end{equation}
where $\mu$ is the dynamic viscosity of the fluid. The pressure difference, $\Delta p'<0$, is calculated between the two sides of the aggregate.

The resistance of the device $R_{tot}$ increases during the growth of the aggregate. The total resistance is the sum of the resistances of the different sections of the device. We observe experimentally that, owing to the small deformation of the PDMS channel, the contaminant typically enters the constriction where it gets trapped. Therefore, the resistance of this section of the microfluidic device becomes $\tilde{R}_{c}> R_c$. The hydraulic resistance of the microchannel increases as the aggregate builds up. The total resistance is time dependent:
\begin{equation} \label{resistance}
R_{tot}(t)=2\,R_{res}+\tilde{R}_{c}+R\,\left(\frac{L-\ell(t)}{L}\right)+\frac{\mu\,\ell(t)}{k\,w\,h}.
\end{equation}
where ${R}_{res}$ denotes the resistance of the reservoir and ${R}$ is the resistance of the channel of length $L$ in the absence of aggregate (open channel). Using Eqs. (\ref{evolution_l}) and (\ref{resistance}), we obtain the evolution equation for the length of the aggregate $\ell(t)$:
\begin{equation} \label{equation_l}
\frac{\rm{d}\ell(t)}{\rm{d} t}=\frac{\alpha\,\Delta p}{\phi\,h\,w} \left[{2\,R_{res}+\tilde{R}_{c}+R+\ell(t)\left(\frac{\mu}{k}\,\frac{1}{w\,h}-\frac{R}{L}\right)}\right]^{-1}.
\end{equation}
Equation (\ref{equation_l}) can be solved analytically to obtain the time evolution:
\begin{equation} \label{solution_1} \displaystyle
\ell(t)=\frac{2\,R_{res}+\tilde{R}_{c}+R}{\dfrac{\mu}{h\,w\,k}-\dfrac{R}{L}}\left[\sqrt{1+\frac{2\,t\,\alpha\,\Delta p\left(\dfrac{\mu}{h\,w\,k}-\dfrac{R}{L}\right)}{\phi\,h\,w\,{\left(2\,R_{res}+\tilde{R}_{c}+R\right)}^2}}-1\right].
\end{equation}
The resistance of the device can then be calculated at any time using Eq. (\ref{resistance}) and the flow rate is given by $Q(t)=\Delta p/R_{tot}(t)$.

The above equation may be simplified depending on the geometry of the microfluidic device. We have measured experimentally the hydraulic resistance of the different sections of the microfluidic device: $R_{res}=5.4\times 10^{11}\,{\rm Pa.s.m^{-3}}$, $R=1.2\times 10^{14}\,{\rm Pa.s.m^{-3}}$ (for $w=40\,\mu{\rm m}$) and $R=8.0\times 10^{13}\,{\rm Pa.s.m^{-3}}$ (for $w=60\,\mu{\rm m}$), $R_c=1.8\times 10^{13}\,{\rm Pa.s.m^{-3}}$. Although this latter resistance increases when the constriction gets clogged, its value remains negligible compared to the hydraulic resistance of the channel with an aggregate, which is about $R_{agg} \sim 10^{17}\,{\rm Pa.s.m^{-3}}$ (for an aggregate of typical length $\ell=100\,\mu{\rm m}$). The aggregate length finally reads
\begin{equation} \label{solution_2} \displaystyle
\ell(t)=\left[\frac{2\,\alpha\,\Delta p\,k}{\mu\,\phi}\,t\right ]^{1/2}.
\end{equation}
 Depending on the hydraulic resistance of the different parts of the microfluidic device and the permeability $k$, the full expression in Eq. (\ref{solution_1}) may be required. The simplified Eq. (\ref{solution_2} can be used when the permeability of the aggregate is small, and at sufficiently long time, \textit{i.e.}, its hydraulic resistance is large.

\subsection{Comparison to experimental measurements}

The analytical expression in Eq. (\ref{solution_2}) captures the experimental results presented in Sec. II.C. Indeed, the modeled length of the aggregate does not depend on the width or the height of the channel, in agreement with the observations in Fig. \ref{syst_exp}(c). This effect comes directly from Darcy's law that states that the rate of particles addition to the aggregate is ${\Delta p'}/{\ell(t)}=-{\mu\,Q}/(k\,h\,w)$. Since the flow rate varies linearly with the dimensions of the cross section, the length of the aggregate does not depend on the width of the microchannel, whereas its volume does.

\begin{figure}
  \begin{center}
\includegraphics[width=0.55\textwidth]{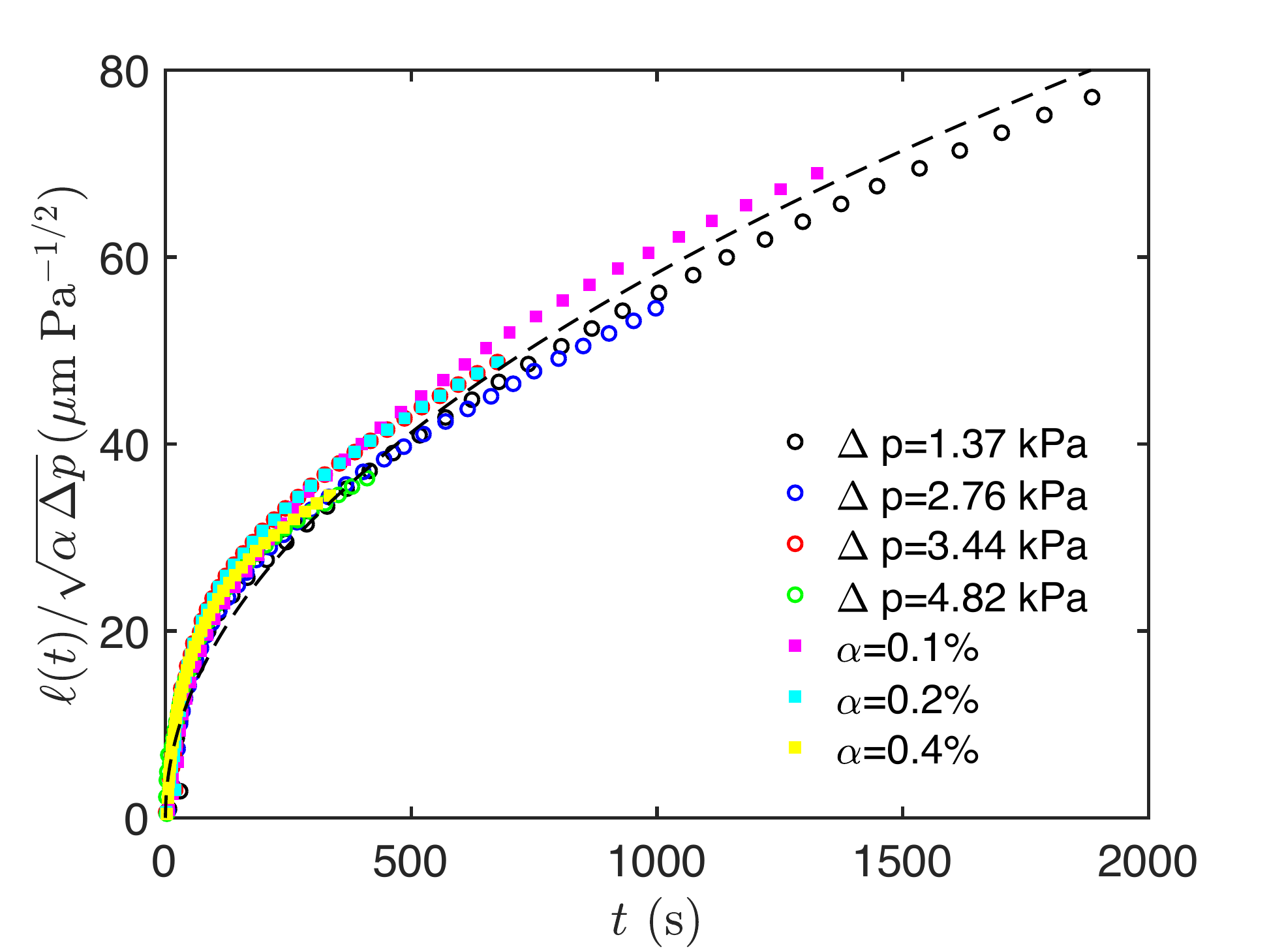}
    \caption{Evolution of $\ell(t)/\sqrt{\alpha\,\Delta p}$ for varying concentrations and pressure in a channel of width $w=40\,{\mu{\rm m}}$. The hollow circles corresponds to $\alpha=0.2\%$ and $\Delta p=13.8,\,27.6,\,34.5,\,48.2{\rm kPa}$ and the filled squares are obtained for $\Delta p=34\,{\rm kPa}$ and varying concentrations $\alpha=0.1\%,\,0.2\%, \,0.4\%$. The dash-dotted line is the expression (\ref{solution_2}) obtained with $\phi=0.70$ and $k$ obtained from the Carman-Kozeny relation.}
    \label{exp_rescaled}
  \end{center}
\end{figure}

In Fig. \ref{exp_rescaled}, we reported the time evolution of the rescaled length, $\ell(t)/\sqrt{\alpha\,\Delta p}$, varying the pressure difference $\Delta p$ and the particle concentration $\alpha$. We observe that all the experimental data collapse well on a master curve. The theoretical curve in this figure is obtained using Eq. (\ref{solution_2}) and fitting the compacity of the aggregate $\phi$. The permeability $k$ of the aggregate is estimated using the Kozeny-Carman relation \cite{bear2013dynamics}:
\begin{equation}
k=\frac{d^2\,(1-\phi)^3}{A\,\phi^2},
\end{equation}
where $A$ is a constant usually taken equal to 180. Using this expression for $k$, the fitted value of the compacity is $\phi \simeq 0.7 $. Although this value is quite large, it is consistent with previous work. Indeed, different studies have shown that the permeability of an aggregate of particles can not be estimated very accurately using the Kozeny-Carman equation that often overestimates the compacity \cite{tien2013can}. In the following, we shall use the experimental parameters for the hydraulic resistances and the properties of the aggregates with the full Eq. (\ref{solution_1}) to study the situation with $ N $ microchannels in parallel. Combining the time evolution of a single aggregate obtained previously with a stochastic model for the clogging of microchannel, we are going to develop a model to describe the time evolution of the flow rate in $N$ microchannels.

 \section{Competitive growth of aggregates in $N$ parallel channels} 

Microfluidic devices with a large number of parallel channels are commonly used to investigate clogging dynamics at the pore scale or to screen biological samples as in cell sorting. The present approach may also be useful to determine the evolution of the flow rate during fracking. It is often assumed that all the microchannels behave identically, and independently of one another (see, \textit{e.g.} Refs. \cite{wyss2006mechanism,dersoir2015clogging}). The common approach with pressure driven flows is to consider that once a channel is clogged, its hydraulic resistance diverges to infinity and the flow rate becomes zero. The flow rate in all the other open channels is assumed to remain constant and equal to $q$.

\subsection{Principle of the numerical model}

To model the clogging dynamics, we use a stochastic model of clog formation for each microchannel coupled with the growth of the aggregate once the channel is clogged. The dynamics is then solved numerically using a custom-made routine. The geometry considered here is similar to the one presented in Fig. \ref{setup}(a), except that we now have $N\in[2,\,72]$ channels of width $w=40\,\mu{\rm{m}}$ in parallel. As shown in Sauret \textit{et al.} \cite{sauret2014clogging}, the clogging of microchannels by sieving can be modeled by a stochastic process that follows a Poisson distribution. The principle of the model is to consider the flow of a suspension driven by a pressure difference $\Delta p$ in the $N$ parallel microchannels of a device. At each time step ${\rm{d}} t$, we calculate the hydraulic resistance of each microchannels, $R_i$, which can be clogged or open. We can then calculate the flow rate $q_i$ in each of the microchannels and the total flow rate $Q = \sum_{i = 1}^N q_i $ in the device. This total flow rate is compared to the direct calculation of the flow rate in the device, given by $ Q = \Delta p / R_{tot} $ to ensure that each channel is accounted for correctly.

During a time step ${\rm{d}} t$, the probability of clogging an open microchannel is equal to $ p_i ({\rm {d}} t) = c \, Q_i \ {\rm {d}} t $ \cite{sauret2014clogging}. At each time step, the algorithm accounts for three possibilities for each microchannel: (i) If it was already clogged at the previous time step, the length of the aggregate is updated, (ii) if it was open and no contaminant enters the channel during $ {\rm {d}} t $, the channel remains open, and (iii) if it was open at the previous time step and the stochastic model predicts clogging, we initialize the length of the aggregate which will then grow in the following time steps. Once this analysis is initialized, steps (i) to (iii) are repeated until all channels are clogged. For the range of parameters used in our numerical simulations and the time over which the calculation is performed, the length of the aggregate in any microchannel remains smaller than the total length of the channel, $\ell(t) < L$.

\subsection{Negligible hydraulic resistance of the reservoir: numerical simulation and theoretical modeling}

We first consider that the hydraulic resistance of the reservoir is negligible compared to the hydraulic resistance of the parallel microchannels, \textit{i.e.}, $R_{res} \ll R/N$. This condition is not fulfilled in the experiment shown in Fig. \ref{exp_40canaux} but is common in the literature \cite{dersoir2015clogging,dersoir2017clogging,liot2017pore}. In each of the following sections, we first present the results of the numerical simulations described in the section IV. A.. Assuming that the flow rate of a clogged channel is null, we can model the system analytically. Numerical and analytical results are then compared.

\subsubsection{Mean time interval between two clogging events}
\begin{figure}
  \begin{center}
\subfigure[]{\includegraphics[width=7cm]{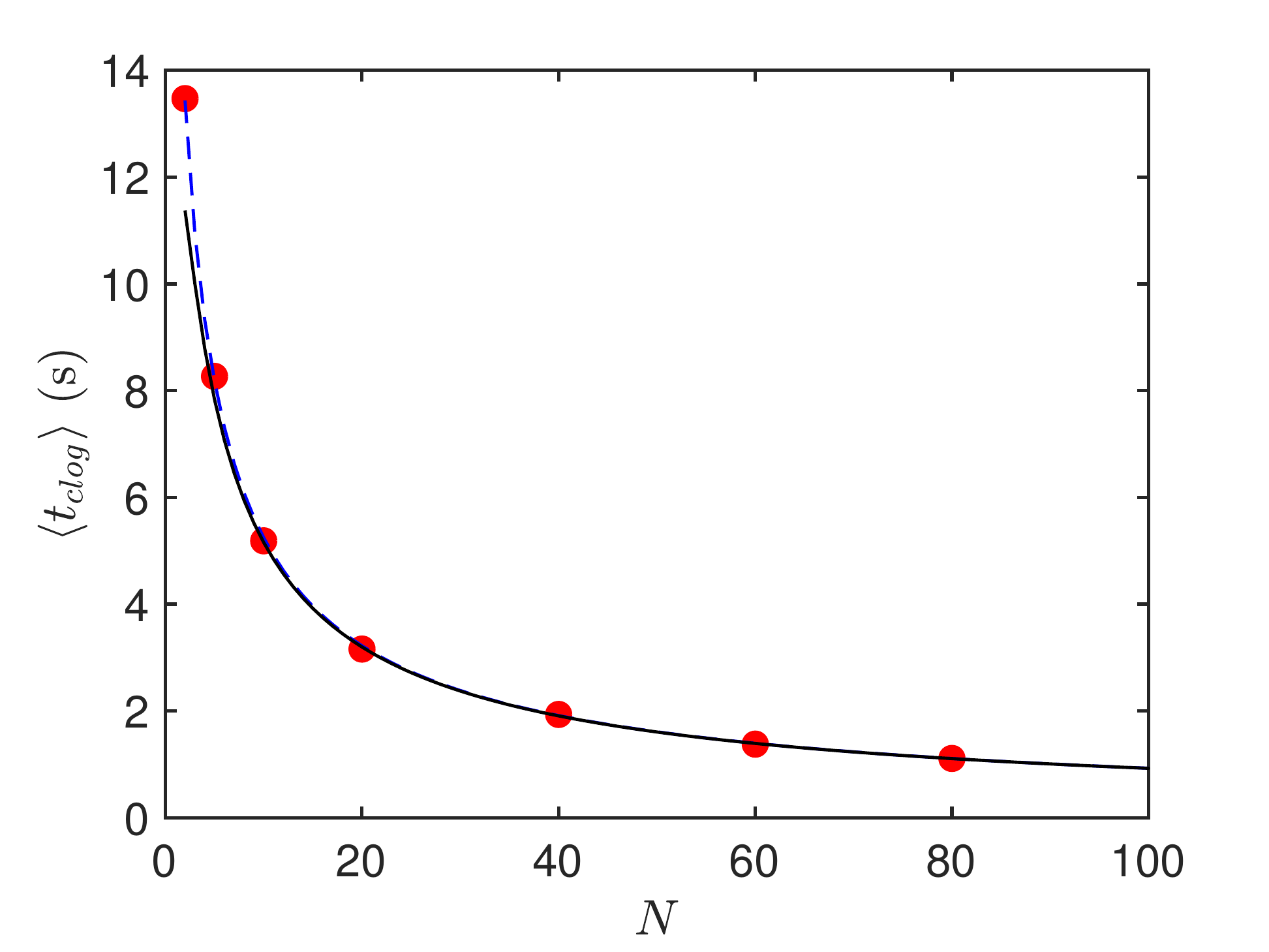}}
\subfigure[]{\includegraphics[width=7cm]{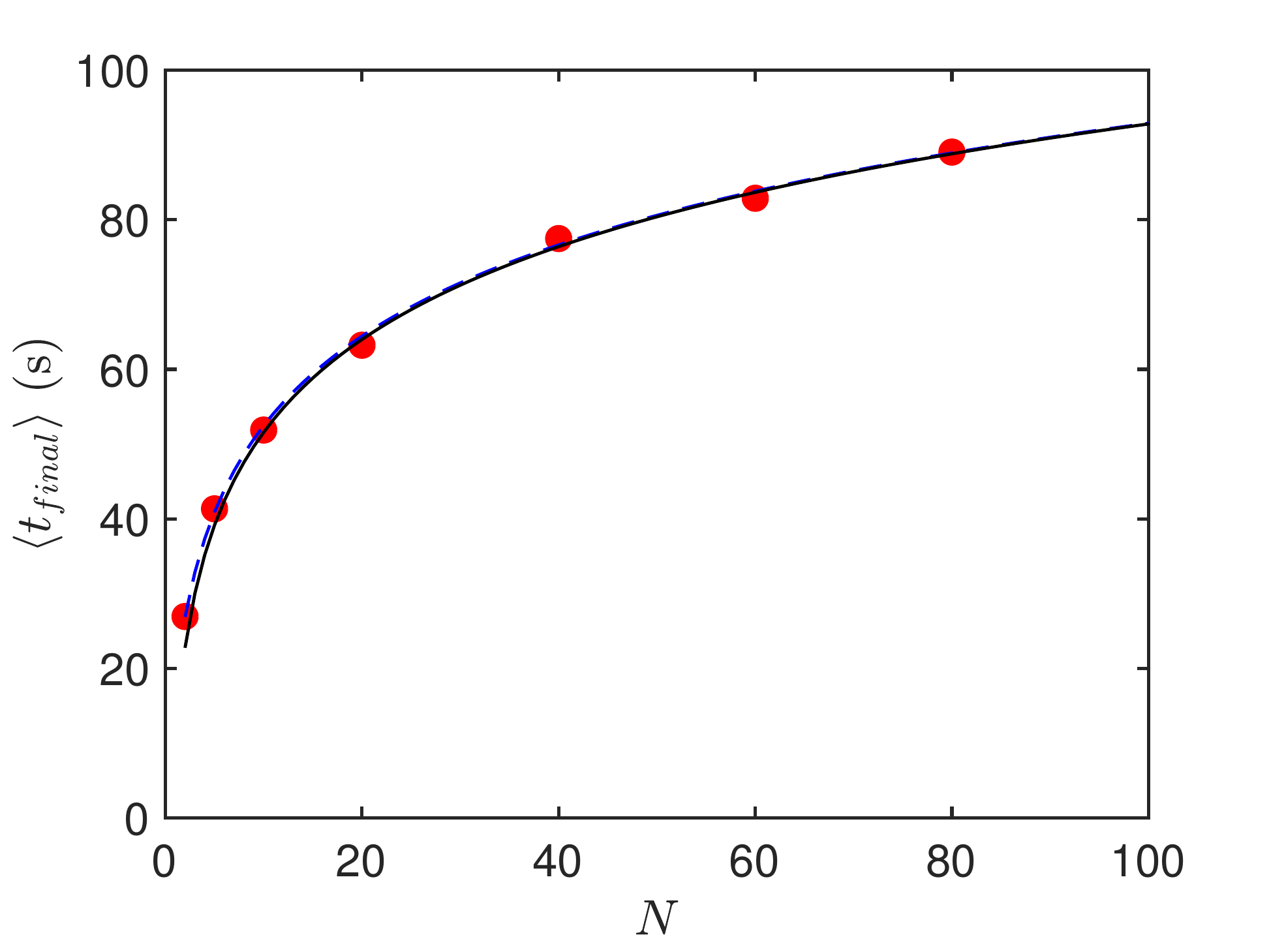}}
    \caption{(a) Mean time interval between two clogging events $\langle t_{clog}\rangle$ for varying number of channel $N$. The circles are the results of the numerical model, the blue dotted line is the analytical solution given in Eq. (\ref{clog_theory_APL_1}) and the continuous black line is the simplified solution (\ref{clog_theory_APL_EV}). (b) Mean clogging time for the entire clogging of the microfluidic device $\langle t_{final}\rangle$ for varying number of channel $N$. The parameters are $\Delta p=13.8\,{\rm kPa}$, $\alpha=0.2\%$, $c=5.6\times10^8 \,{\rm m^{-3}}$, and the values of the hydraulic resistance and compacity of the clog are the same as in the previous section.}
    \label{timeclog}
  \end{center}
\end{figure}

We start describing the average time between two clogging events and the total time during which the microfluidic device can function before clogging by using simple arguments. Numerical simulations are presented in Fig. \ref{timeclog}(a)-(b) for a varying number of parallel microchannels $N$. For each set of parameters, 1000 numerical experiments are averaged. The mean time interval between two clogging events $\langle t_{clog}\rangle$ decreases when increasing $N$ because the volume of suspensions filtered per unit time increases. Also, the clogging time, which corresponds to the complete clogging of the microfluidic device $\langle t_{final}\rangle$ increases with $N$ leading to a longer lifetime of the device.

We make the simple assumption that, upon clogging, the hydraulic resistance of a channel is equal to $ R_i = + \infty $ with flow rate $ q_i \to 0 $ instantaneously. Other options, like those investigated by Hermia \cite{Hermia} are possible (see Section \ref{casc}). The time interval between the clogging of the $(i-1)$-th and the $ i $-th channel is given by $ t_{clog, i} = 1 / [c \, Q (i)] $ where $ Q (i) = \Delta p / R_{tot} (i) $ is the total flow rate in the microfluidic system at that time. Within this framework, the mean time interval between two clogging events, $\langle t_ {clog} \rangle$, is
\begin{equation}\label{clog_theory_APL_1}
\langle t_{\rm clog}\rangle=\frac{1}{N}\sum_{i=0}^{N-1}\frac{R/(N-i)}{c\,\Delta p}=\frac{1}{N\,q\,c}\sum_{j=1}^{N}\frac{1}{j},
\end{equation}
where $q=\Delta p/R$ is the flow rate in a single microchannel. As a result $ \langle t \rangle ={1}/{(q\,c)}$ is the mean clogging time in a single microchannel. Since $H_N=\sum_{j=1}^N j^{-1} \simeq {\rm ln}\, N+ \gamma$ is the harmonic number of order $N (\gg 1)$, with $\gamma \simeq 0.577 $ being the Euler's gamma constant, Eq. (\ref{clog_theory_APL_1}) can be written
\begin{equation}\label{clog_theory_APL_EV}
\langle t_{\rm clog}\rangle=\langle t \rangle  \frac{H_N}{N}.
\end{equation}
The time to clog entirely the microfluidic device, \textit{i.e.}, the $N$ microchannels, is 
\begin{equation} \label{tfinal}
\langle t_{\rm final}\rangle=N\,\langle t_{\rm clog}\rangle = \langle t \rangle \,H_N,\end{equation}
and both $\langle t_{\rm clog}\rangle$ and $\langle t_{\rm final}\rangle$ are plotted in Fig. \ref{timeclog}(b). In the present situation where the structure of the aggregate is imposed only by steric effects, thus leading to low porosity, the results of our numerical model and the comparison with the analytical modeling show that we can assume at first order that the flow rate becomes negligible in a clogged channel shortly after the sieving occurred. 

\subsubsection{Distribution of clogging times} \label{distributin_EV}
\begin{figure}
  \begin{center}
 \subfigure[]{\includegraphics[width=7.5cm]{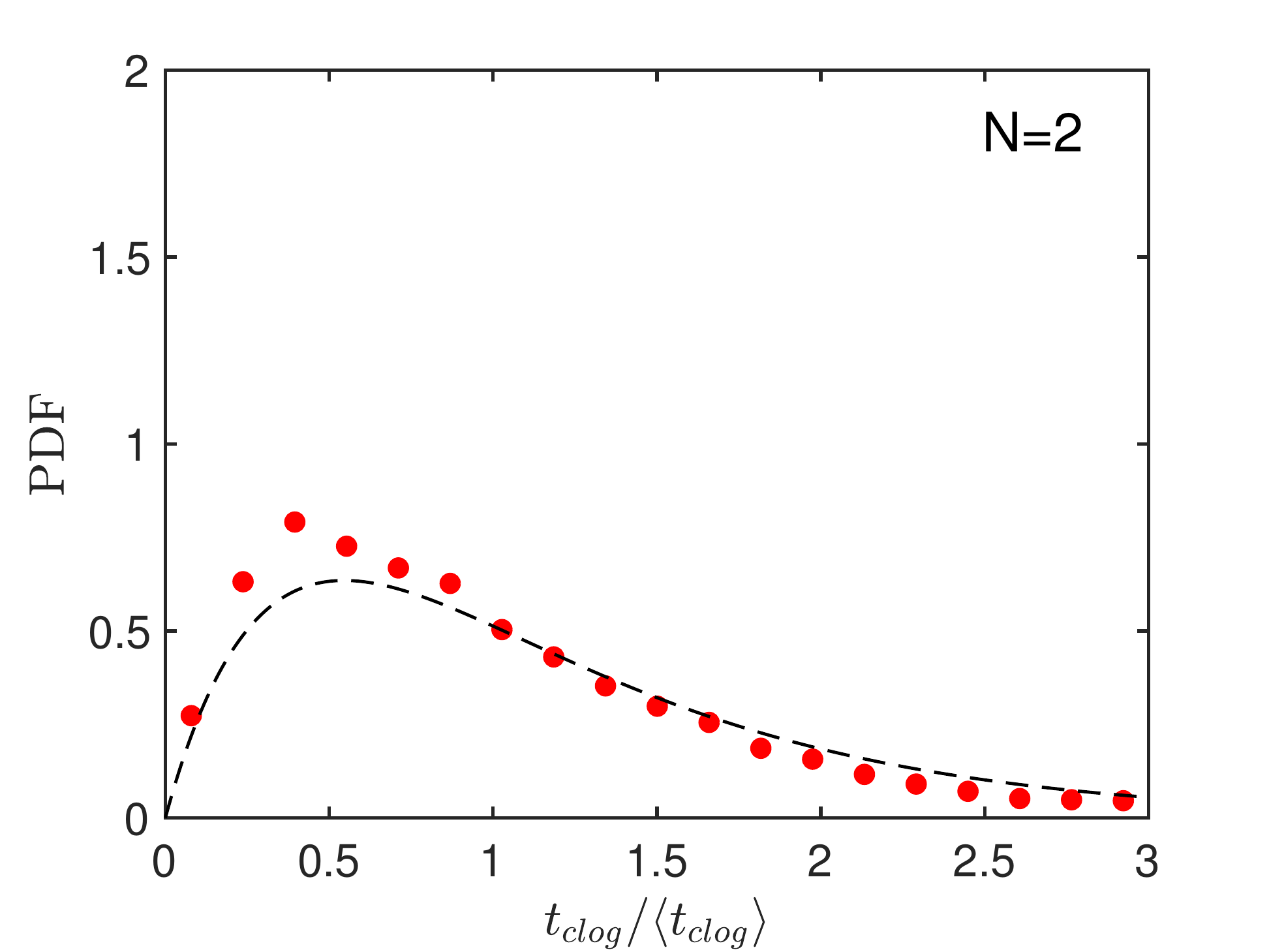}}
  \subfigure[]{\includegraphics[width=7.5cm]{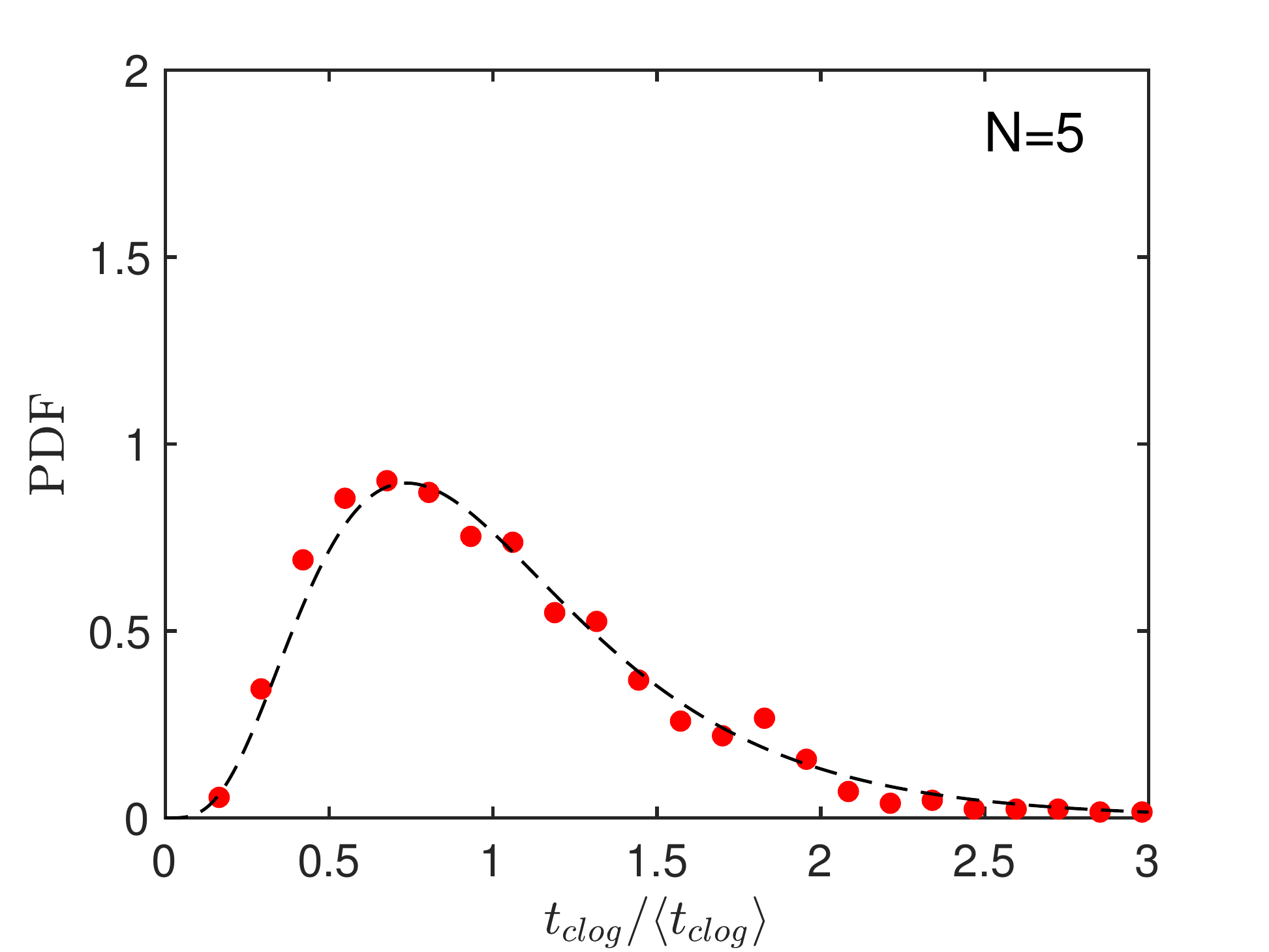}} \\
   \subfigure[]{\includegraphics[width=7.5cm]{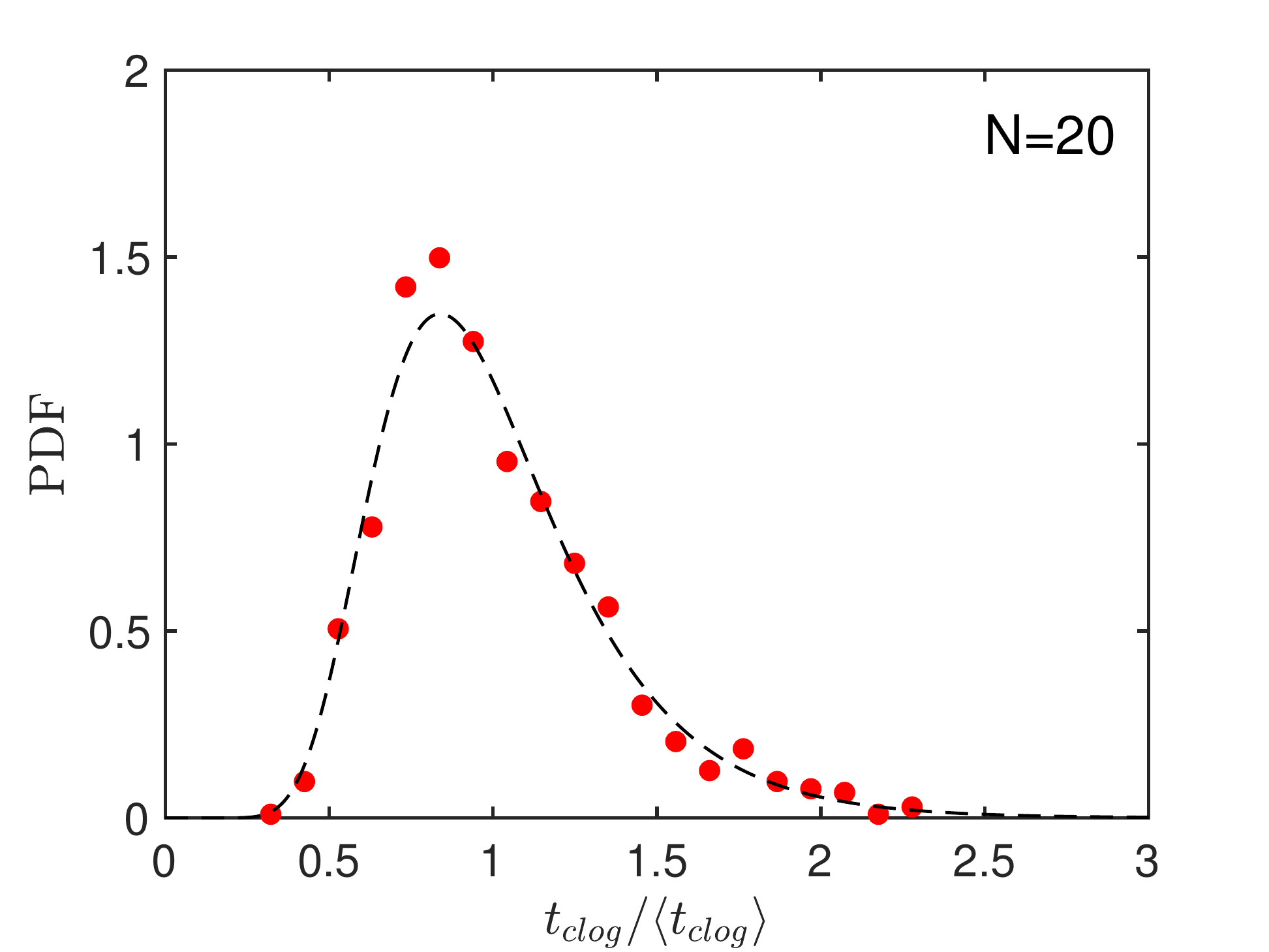}}
  \subfigure[]{\includegraphics[width=7.5cm]{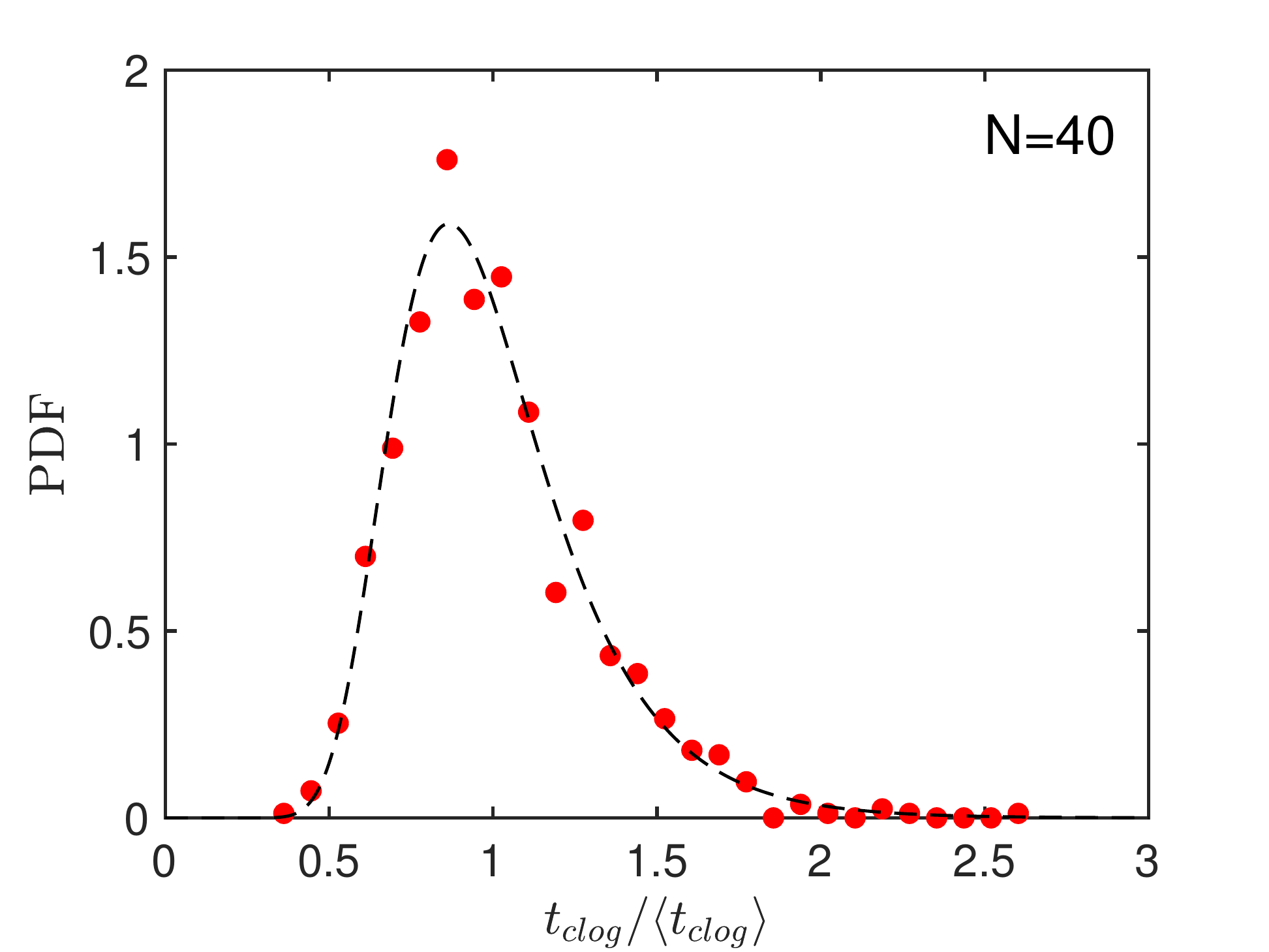}} 
    \caption{Dimensionless distribution of the clogging times for microfluidic devices made of $N=2$, $N=5$, $N=20$, $N=40$. The experimental parameters are the same than in Fig. \ref{timeclog}. The dotted line shows the expression (\ref{ptclogx}).}
    \label{gamma}
  \end{center}
\end{figure}

A more comprehensive description of the problem requires the knowledge of the entire clogging time distribution. One sees in Fig. \ref{gamma} that the probability density function (PDF) of the rescaled clogging time $t_{\rm final}/\langle t_{\rm final} \rangle$ has a skewed bell-shape, and is more peaked for a larger number of parallel microchannels $N$.

For $N$ independent channels in parallel, the probability that each channel clogs is described by a pure random process. The flow rate per channel $q$ is constant, independent of time, and independent of the number of channels previously clogged. Therefore, the usual construction of the Poisson distribution holds. The probability of clogging during the time interval ${\rm d} t$ is extensive to ${\rm d} t$ and is equal to ${\rm d} t/\langle t\rangle$. The probability that a given channel has not clogged at time $T$ is
\begin{equation}
P(t>T)=\left(1-{\rm d} t/\langle t\rangle\right)^{T/{\rm d} t}\xrightarrow[{\rm d} t\to 0]{} e^{-T/\langle t\rangle}.
\end{equation}
If $p(t) \,{\rm d}t$ is the probability that the channel clogs between the time $t$ and $t+{\rm d}t$, we have $P(t>T)=\int_T^\infty p(t){\rm d}t$, and $p(t)=-{\rm d}P(T)/{\rm d}T\vert_{T=t}$, giving
\begin{equation}
p(t)=\frac{e^{-t/\langle t\rangle}}{\langle t\rangle}.
\label{poissoneq}
\end{equation}

The probability density function $p(t_{\rm final})$ of the final clogging time, namely the probability $p(t_{\rm final}){\rm d}t_{\rm final}$ that the microfluidic device of $N$ channels in parallel clogs between the time $t_{\rm final}$ and $t_{\rm final}+{\rm d}t_{\rm final}$ is the product of two probabilities, namely: $(i)$ the probability that, among $N$ channels, $N-1$ have already clogged at  $t_{\rm final}$ and $(ii)$, the probability that the $N^{\rm th}$, \textit{i.e.}, the last channel clogs between $t_{\rm final}$ and $t_{\rm final}+{\rm d}t_{\rm final}$. Clogging of the $N-1$ channels can occur at random among the $N$ channels available in the device, so that this product of probabilities should be weighed by a factor $N!/(N-1)! 1!=N$. 

Since $P(t<t_{\rm final})=\int_0^{t_{\rm final}}p(t){\rm d}t=1-e^{-t_{\rm final}/\langle t\rangle}$, the probability $(i)$ is $\left(1-e^{-t_{\rm final}/\langle t\rangle}\right)^{N-1}$ whereas the probability $(ii)$ is given by Eq. \eqref{poissoneq}, and we have
\begin{eqnarray}
\begin{aligned}
p(t_{\rm final}){\rm d}t_{\rm final}&=N\times\left(1-e^{-t_{\rm final}/\langle t\rangle}\right)^{N-1}\,{\times} \frac{e^{-t_{\rm final}/\langle t\rangle}}{\langle t\rangle}{\rm d}t_{\rm final}\\
{\rm that\,\, is}\quad p(t_{\rm final})&=\frac{N}{\langle t\rangle}\left(1-e^{-t_{\rm final}/\langle t\rangle}\right)^{N-1}e^{-t_{\rm final}/\langle t\rangle},
\end{aligned} \label{ptclog}
\end{eqnarray}
which is such that $\int_0^\infty p(t_{\rm final}){\rm d}t_{\rm final}=1$. This distribution has a mean given by $\langle t_{\rm final} \rangle$ defined by Eq. (\ref{tfinal}) and thus
\begin{align}
\quad p(x=t/\langle t_{\rm final}\rangle)=N\,H_N\left(1-e^{-x\,H_N}\right)^{N-1}e^{-x\,H_N},\label{ptclogx}
\end{align}
and it is reported in Figs. \ref{gamma}(a) and \ref{gamma}(d), capturing well the distribution of the clogging time when the hydraulic resistance of the reservoir is negligible. 

\subsubsection{Clogging cascade} \label{casc}
 \begin{figure}
  \begin{center}
\subfigure[]{\includegraphics[width=7cm]{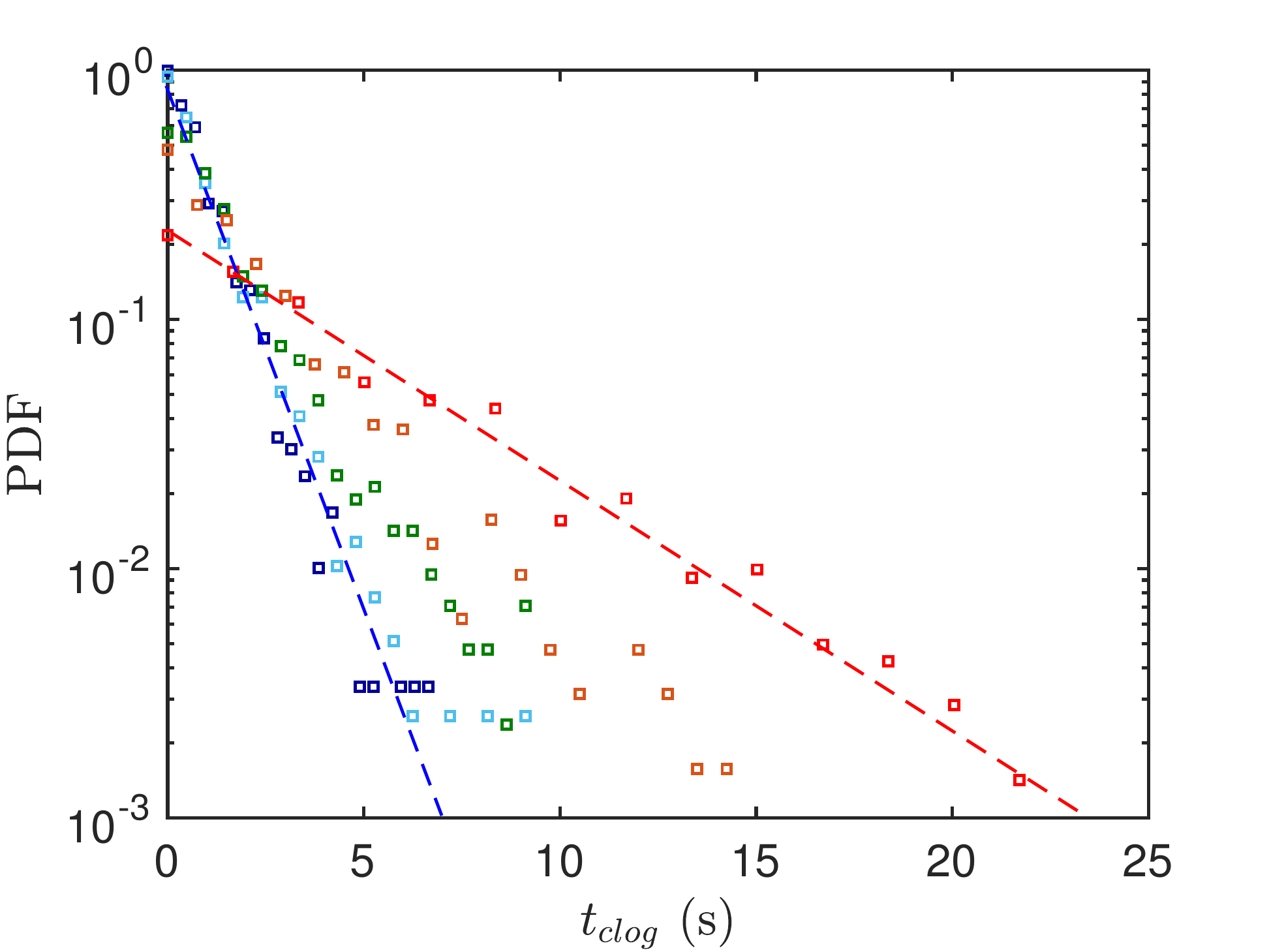}} \quad
\subfigure[]{\includegraphics[width=7cm]{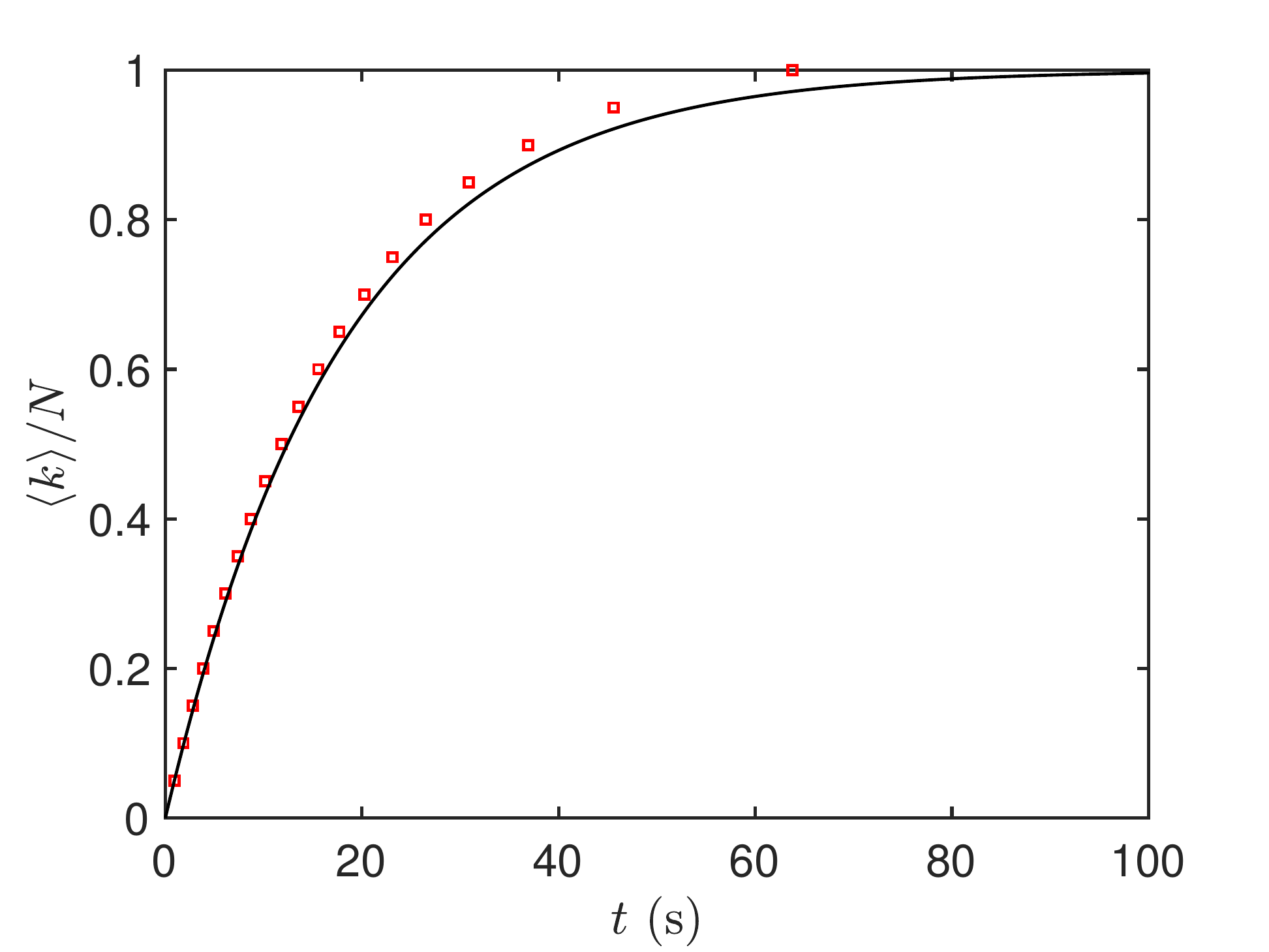}}
    \caption{(a) Distribution of clogging time intervals $t_{clog}$ for $N=20$ channels and the same experimental parameters as in Fig. \ref{timeclog}. The dark blue squares correspond to the clogging of the first channel, and the red squares to the clogging of the $17$-th channel (the other points corresponds to the clogging of the $5$-th, $9$-th and $13$-th). The dotted lines are the best fit assuming a Poisson distribution. (b) Number of clogged channels as a function of time for a device consisting of $N=20$ microchannels in parallel and the same parameters as in Fig. \ref{timeclog}.}
    \label{poisson}
  \end{center}
\end{figure}

\begin{figure}
  \begin{center}
 \subfigure[]{\includegraphics[width=7.2cm]{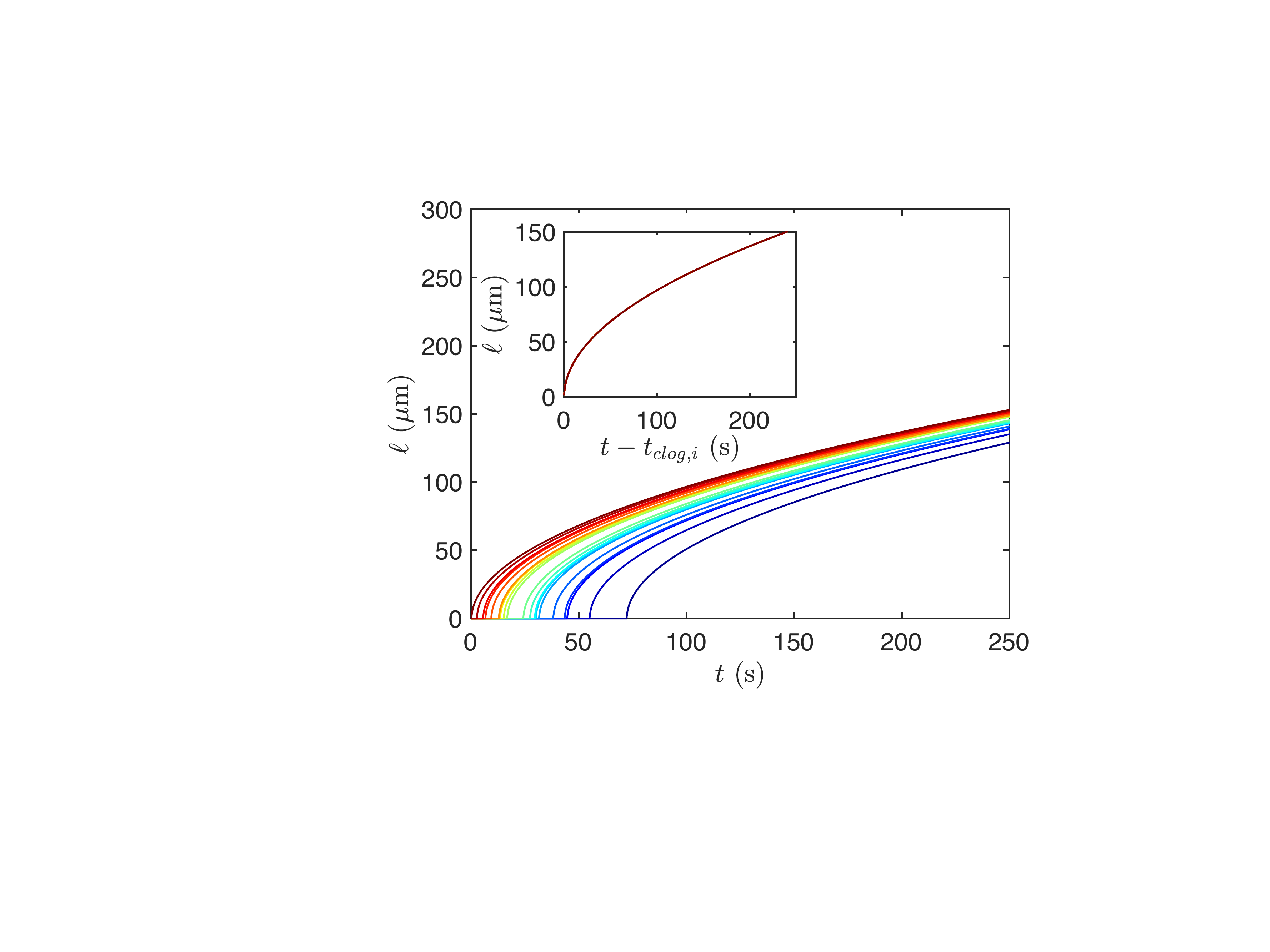}} \quad
 \subfigure[]{ \includegraphics[width=7.0cm]{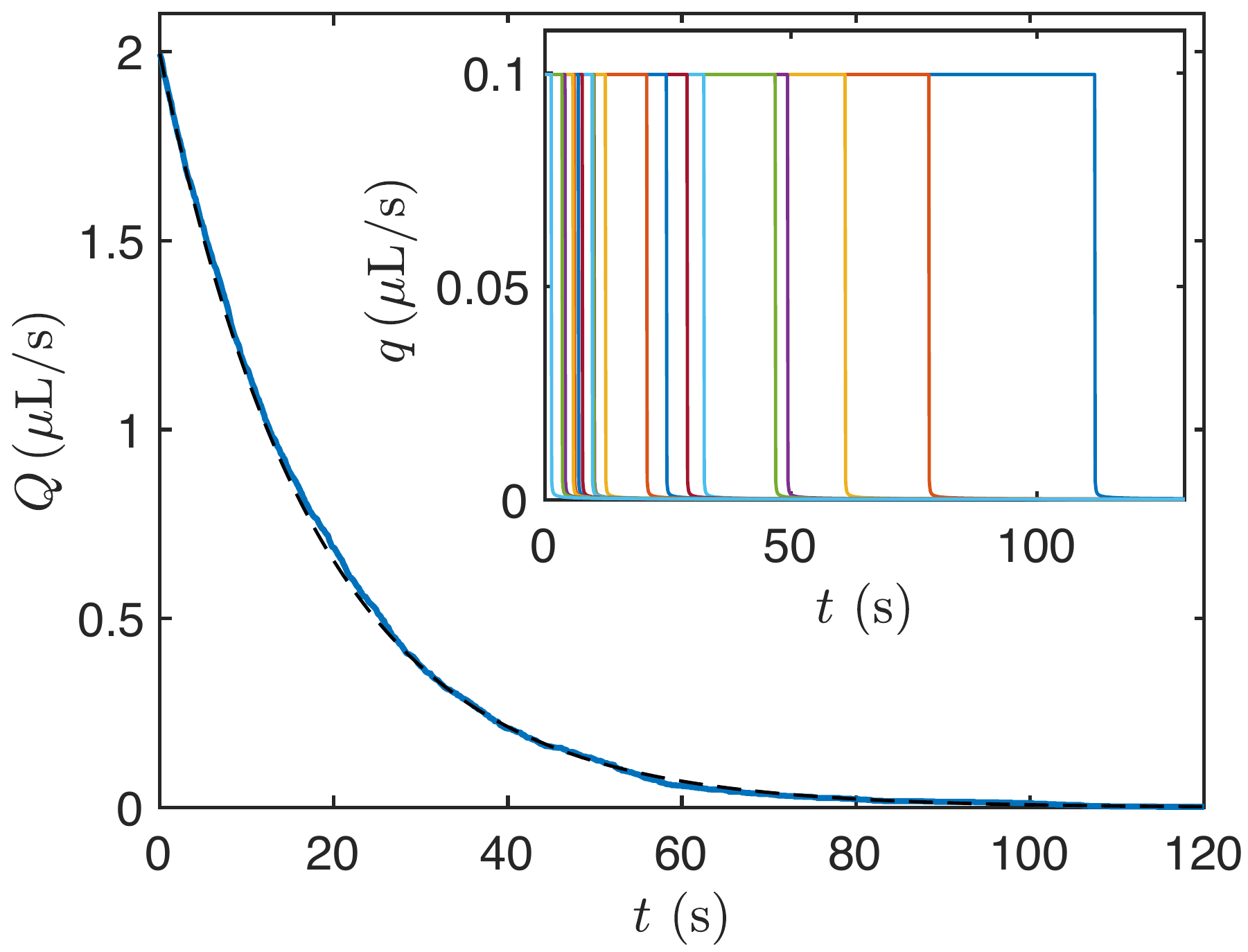}}
    \caption{(a) Time evolution of the aggregate length for $N=20$ channels. The clogs form from red to blue. Inset: rescaled evolution of the aggregate. (b) Time evolution of the flow rate in the microfluidic device consisting of $N=20$ microchannel in parallel averaged over 1000 numerical experiments. The dotted line is the analytical expression (\ref{flow}). Inset: evolution of the flow rate in each individual microchannel obtained in a numerical experiment. In all figures the parameters are the same as in Fig. \ref{timeclog}.}
    \label{lengthagg}
  \end{center}
\end{figure}

The numerical simulations show that the distributions of clogging time intervals for each channel follow a Poisson distribution, with an increasing slope for decreasing number of open channels (Fig. \ref{poisson}(a)). Indeed, the probability to clog a microchannel is proportional to the number of channels in which the suspension and the contaminants flow. Thus, the slope of the Poisson distribution is equal to $1 / (Q \, c) = 1 / [(N-k) \, q \, c)] $ where $ N-k$ is the number of microchannels that remain open, decaying in time, as seen in Fig. \ref{poisson}(b). Such observations were also made using different clogging processes such as aggregation \cite{wyss2006mechanism,van2018cooperative}. After an initial period of frequent clogging events, the final clog formation is slower, as seen from the evolution of the aggregate length in Fig. \ref{lengthagg}(a).

Extending the approach presented in the previous section, we now calculate the most general distribution describing the problem, that is the probability $p(k\vert t)$ of having $k$ channels clogged at time $t$ under the same assumption of constant flow rate per independent channels. As above, the Binomial counting of $k$ channels among $N$, with the individual clogging probability $1-e^{-t/\langle t\rangle}$, where $\langle t\rangle=1/(q\,c)$, leads to 
\begin{equation}
p(k\vert t)=\frac{N!}{(N-k)!k!}\left(1 -e^{-t/\langle t\rangle}\right)^k e^{-(N-k)\,t/\langle t\rangle},\label{binom}
\end{equation}
where $\sum_{k=0}^N\,p(k\vert t)=1$. The net clogging time distribution of the array of $N$ microchannels given in Eq. \eqref{ptclog} is the clogging time distribution of the $N^{\rm th}$ channel, \textit{i.e.}, setting $k=N-1$ in Eq. \eqref{binom}. Therefore, the mean number of clogged channels $\langle k\rangle$ at time $t$ is
\begin{equation} \label{EV_channel}
\langle k\rangle=\sum_{k=0}^N kp(k\vert t)=N\left(1 -e^{-t/\langle t\rangle}\right),
\end{equation}
a relation which captures very well the time evolution of the fraction of clogged channel in Fig. \ref{poisson}(b).

 The above result is also useful to compute the temporal evolution of the (mean) net flow rate passing through the array of microchannels. Since the flow rate $q$ in each opened channel is constant in time (see inset of Fig. \ref{lengthagg}(b)), the total flow rate $Q$ is 
\begin{equation}
Q(t)=\langle (N-k)q\rangle,\quad{\rm or}\quad Q=N\,q\left(1-\frac{\langle k\rangle}{N}\right)
\end{equation}
where $\langle k\rangle$ is the mean number of clogged channels at time $t$ given in Eq. (\ref{EV_channel}). We therefore obtain
\begin{equation}
Q(t)=N\,q\,e^{-t/\langle t\rangle}.
\label{flow}
\end{equation}
The global flow rate decays exponentially (Fig. \ref{lengthagg}(b)), as a consequence of a simple Poisson process, on a timescale given by the clogging timescale of a single channel $\langle t\rangle=1/(q\,c)$. This classical result familiar in filtration processes (see e.g. Hermia \cite{Hermia}) is associated with the fact that we have canceled the flow rate in each channel once clogged, the flow rate $Q(t)$ being thus extensive to the number of remaining unclogged channels at time $t$. The exponential decay is reminiscent of Beer-Lambert's law of attenuation by surface blocking, time $t$ standing for the penetration depth in the medium charged with absorbing particles in this geometrical interpretation.

Other limits can be considered, however, as in the so-called intermediate blocking regime of filtration for which the interstitial fluid can flow through the filtration cake \cite{Hermia} and where $Q(t)=N\,q/(1+t/\langle t\rangle)$, in our notations.

Note finally that the above result in Eq. \eqref{flow} is also a way of understanding the global mean clogging time $\langle t_{\rm final}\rangle=\langle t\rangle H_N\sim \langle t\rangle\,\ln N$: The net, non-zero flow rate $Q$ cannot be smaller than $q$ and is actually equal to $q$ when the $N^{\rm th}$ channel is about to clog. The time it takes to reach this ultimate condition is obtained by setting $Q=q$ in Eq. \eqref{flow}, leading to
\begin{equation}
\langle t_{\rm clog}\rangle\approx \langle t\rangle\,\ln N,
\end{equation}
as expected from Eq. \eqref{tfinal}.

\subsection{Influence of the reservoir}
 \begin{figure}
\begin{center}
\subfigure[]{\includegraphics[width=7cm]{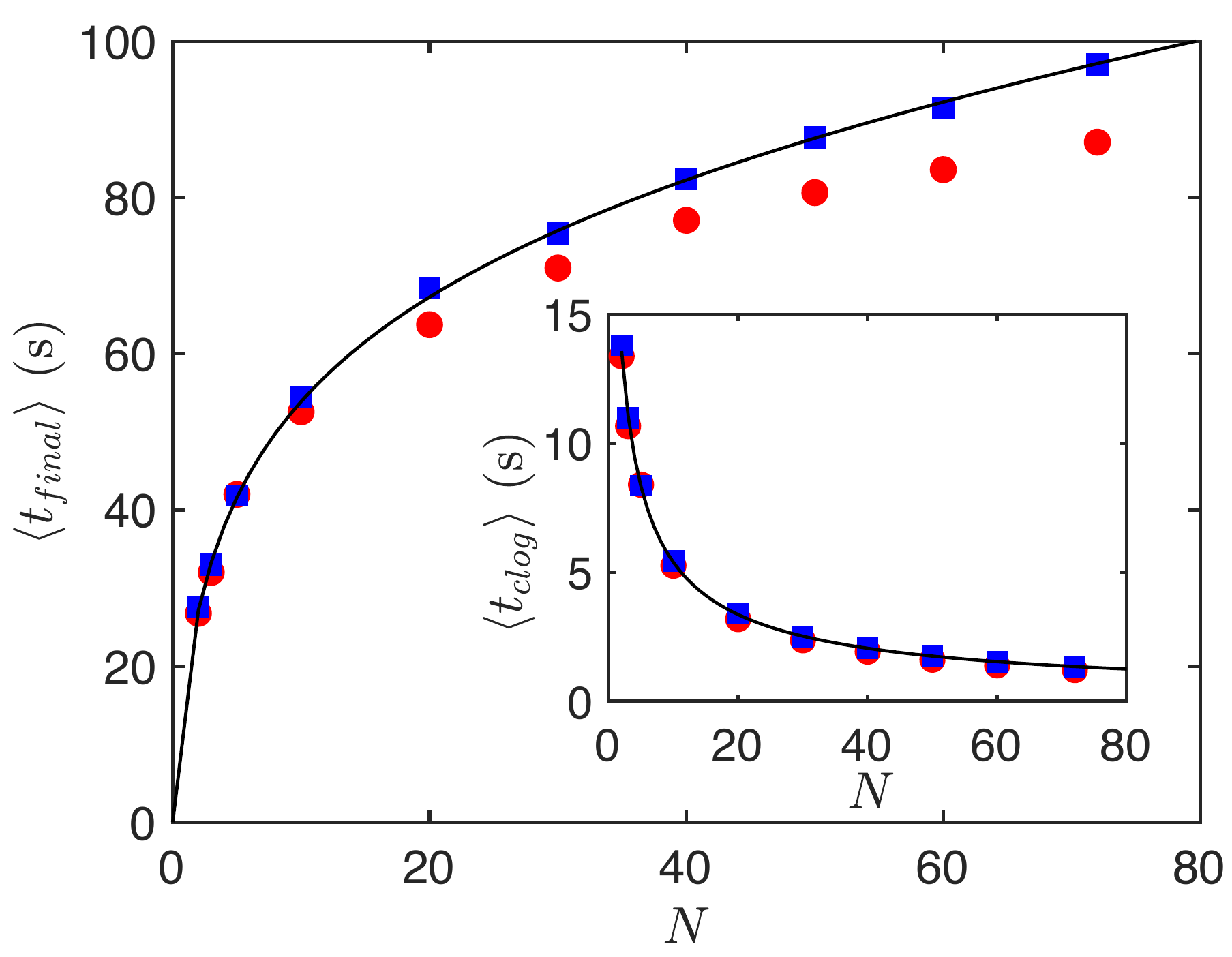}} \quad
\subfigure[]{\includegraphics[width=7.4cm]{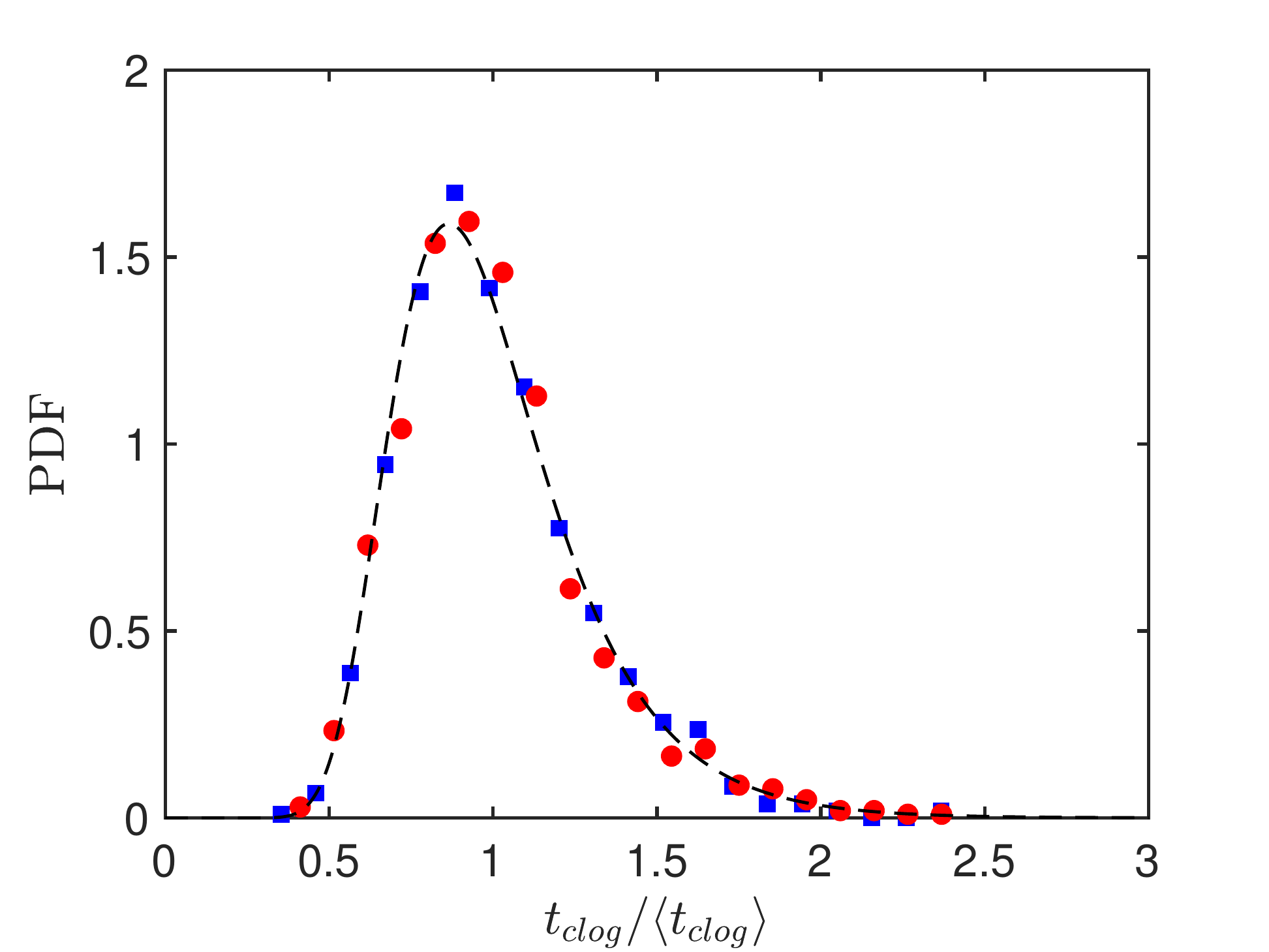}}
    \caption{(a) Mean clogging time for the entire clogging of the microfluidic device $\langle t_{final}\rangle$ for varying number of channel $N$ when the hydraulic resistance of the reservoir is neglected (red circles) or not neglected (blue squares). The parameters are $\Delta p=13.8\,{\rm kPa}$, $\alpha=0.2\%$, $c=5.6\times10^8 \,{\rm m^{-3}}$ and the continuous line is the Eq. (\ref{eq_res_2}). Inset: Mean time interval between two clogging events $\langle t_{clog}\rangle$ in the same situation. The continuous line is the Eq. (\ref{eq_res_1}). (b) Dimensionless distribution of the clogging times for devices having $N=40$ microchannels in parallel when the hydraulic resistance of the reservoir is neglected (red circles) or not neglected (blue squares). The dotted line shows the expression (\ref{ptclogx}).}
\label{reservoir1}
\end{center}
\end{figure}

In Sec. IV.B, we neglect the influence of the reservoirs on the clogging process, assuming their hydraulic resistance is $R_{res}=0$. This assumption allows us to compare the results of our numerical experiments with the theoretical calculations. In this section, we are considering the experimental system presented in Fig. \ref{exp_40canaux} as an example of situation in which the reservoirs have a non-negligible hydraulic resistance. Experimentally, we determine that the hydraulic resistance of one reservoir is $R_{res}=5.4\times 10^{11}\,{\rm Pa.s.m^{-3}}$ and the resistance of one microchannel, where the sieving occurs, is $R_c=1.38\times 10^{14}\,{\rm Pa.s.m^{-3}}$ (for the microchannel $w=40\,\mu{\rm m}$). The maximum number of parallel microchannels that can be fitted between the two reservoirs is $N=72$. In this configuration, the total hydraulic resistance of the  device is $R_{tot}=2\,R_{res}+R/(N-k)$, where $N-k$ is the number of channels that remain open at time $t$ and we assume that the hydraulic resistance of a clogged channel is infinite. To neglect the hydraulic resistance of the reservoir, the condition $2\,R_{res} \ll R/N$ must be fulfilled, which is not the case with the device presented in Fig. \ref{exp_40canaux} with $N=40$ microchannels in parallel.
    
 We perform numerical simulations accounting for the hydraulic resistance of the reservoirs. We report the mean time interval between two clogging events $\langle t_{clog}\rangle$ and the mean time to clog the microfluidic device entirely $\langle t_{final}\rangle$ in Fig. \ref{reservoir1}(a). We observe a clear discrepancy between the two situations, especially when the number of parallel microchannels increase and the condition $2\,R_{res} \ll R/N$ is no longer fulfilled. We can modify the relation established previously to take into account the hydraulic resistance of the reservoir. We obtain a modified expression for $\langle t_ {clog} \rangle$:
\begin{equation}\label{eq_res_1}
\langle t_{\rm clog}\rangle=\frac{1}{N}\sum_{i=0}^{N-1}\frac{2\,R_{res}+R/(N-i)}{c\,\Delta p},
\end{equation}
and because the time required to clog the $N$ microfluidic channels is equal to $N$ times the average time to clog a channel, we have 
\begin{equation} \label{eq_res_2}
\langle t_{\rm final}\rangle=N\,\langle t_{\rm clog}\rangle = \sum_{i=0}^{N-1}\frac{2\,R_{res}+R/(N-i)}{c\,\Delta p}.
\end{equation}

The results reported in Fig. \ref{reservoir1}(a) (main panel and inset) show that the modified expression captures well the mean lifetime of the microfluidic device, even when the resistance of the reservoir cannot be neglected. We also consider the PDF of the clogging times in Fig. \ref{reservoir1}(b), and we observe that this distribution is not significantly modified at first order.

\begin{figure}
  \begin{center}
 \subfigure[]{\includegraphics[width=7.2cm]{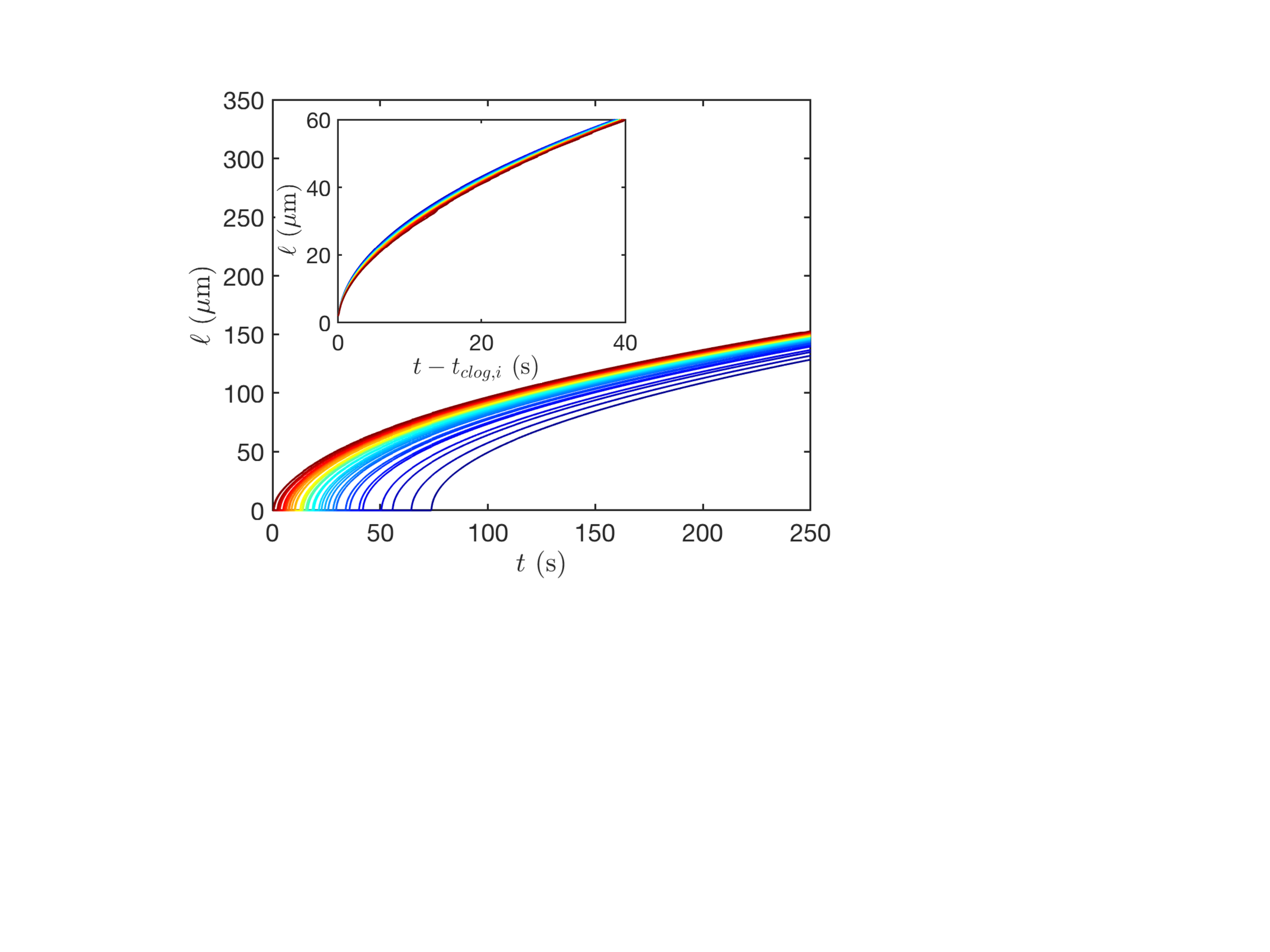}} \quad
 \subfigure[]{ \includegraphics[width=7.0cm]{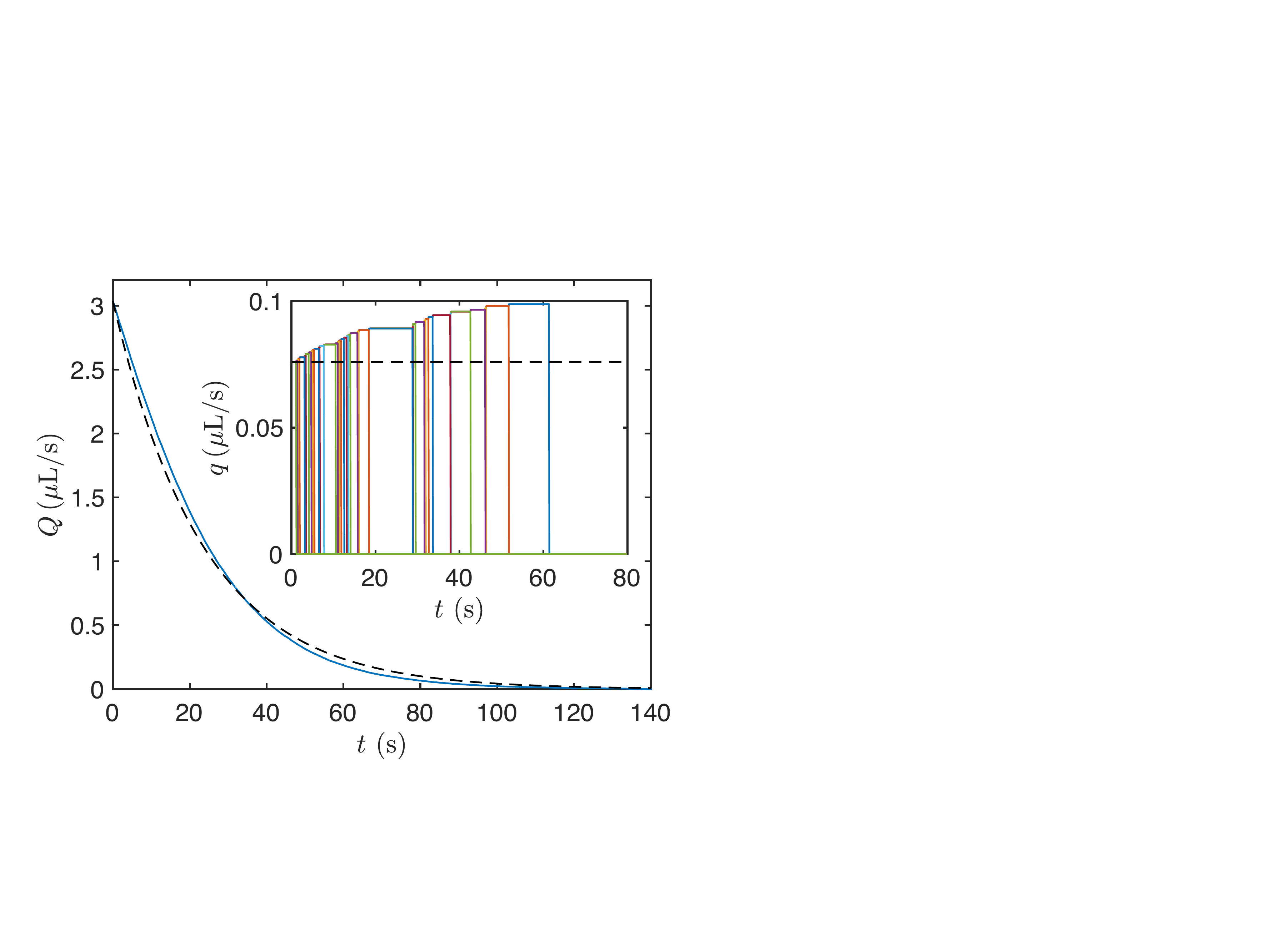}}
    \caption{(a) Time evolution of the aggregate length for $N=40$ channels when considering the presence of the reservoirs. The clogs form from red to blue. Inset: rescaled evolution of the aggregate. (b) Time evolution of the averaged flow rate in the microfluidic device consisting of $N=40$ microchannels. The dotted line is the analytical expression (\ref{flow}). Inset: evolution of the flow rate in each individual microchannel obtained in a numerical experiment.}
     \label{lengthagg2} \end{center}
\end{figure}

However, the hydraulic resistance of the reservoirs modified the growth of the aggregates. When the hydraulic resistance of the reservoir is negligible, the aggregates grow independently. Here, we show that the first aggregates have a slower growth than the following ones [inset in Fig. \ref{lengthagg2}(a)]. This change in growth rate is related to a modification of the flow rate when the channels successively clog. The clog growth determines the total flow rate in the microfluidic device, as presented in Fig. \ref{lengthagg2}(b): Equation (\ref{flow}) accurately captures the global trend but not the time evolution of the flow rate in the microfluidic device. This difference comes from the variation in the flow rate in each individual channel remaining open during the clogging process, as reported in the inset of Fig. \ref{lengthagg2}(b). While initially the flow rate in each channel is well described by neglecting the influence of the reservoir, when more microchannels become clogged, the flow rate in each remaining open channel increases, with a change in flow rate up to 20\% in the situation considered here.

These results show that the condition $ 2 \, R_{res} \ll R / N $ must be valid until all channels are clogged to be able to consider the microchannels as independent. To fullfill this condition, one option is to build a reservoir with a larger height than the rest of the device, as done in different recent studies \cite{dersoir2015clogging,dersoir2017clogging,liot2017pore}.

\section{Conclusions}

In this paper, we provide a description of the time evolution of the clogging dynamics and the flow rate evolution in a microfluidic device comprising $N$ parallel microchannels. To obtain a reliable model, we first obtain theoretically the evolution of the length of the aggregate in a single microchannel and compare it with experimental results. This first result is then used to develop our model.

In summary, we demonstrate that the growth of the aggregate can be described by an analytical solution in the case of a pressure driven flow. The length of the aggregate is proportional to $ \sqrt{\alpha \, \ \Delta p\,t} $, where $ \alpha $ and $ \Delta p $ represent the concentration of the suspension and the pressure difference between the inlet and the outlet of the microfluidic device, respectively. 

We also provide a general expression that can be used when the hydraulic resistance of the different parts of the device cannot be neglected.

In a second time, thanks to the modeling in a single microchannel, we develop a numerical model of the clogging of a microfluidic device that combines a stochastic clogging process and the growth of the resulting aggregates. The results of the numerical simulations are successfully rationalized using analytical modeling when the resistance of the reservoirs is negligible. We predict quantitatively the drop in flow rate in the microfluidic device. Although the clogging process considered here is sieving, the same stochastic description should apply to jamming processes when considering a concentrated suspension and the results should be applicable. 

We illustrate the limits of a model based on the assumption that the flow rate is equal to zero in a microfluidic channel when it is clogged by a particle. The influence of the flow remaining in the microchannel after clogging can become important depending on the porosity and hydraulic resistance of the aggregate. In this case, it may sometimes be necessary to take into account the residual flow rate remaining in each clogged channels. This stochastic model can also be used to optimize the design of microfluidic device used for the detection of cancer cells or used to study clogging by aggregation.

A possible application of the above result for $Q$ in Eq. \eqref{flow} may be found in the context of the fracking industry. It is known that the extraction flow rate of oil from the shale is a decreasing function of time. Several very different scenarios exist to explain this fact, which remains largely elusive \cite{Marder_13,bocquet_16}. 
 So-called proppant particles (sand grains) are dispersed in the pressurized fracking water stream to prevent fractures from closing when the pressure is released. Occasionally, these particles escape, and accumulate in the restricted portions of the fractured channels very much as we have described here. It is therefore possible that the ultimate decrease of the flow rate, which is known to have an exponential-like decay in time \cite{Marder_13}, may result from the present clogging mechanism.

 \section*{Acknowledgements}

The work of A.S. and E.D. is partially supported by a CNRS PICS Grant N$^{\rm o}$ 07242. This research was made possible thanks to the ACS-PRF 55845-ND9 grant from the American Chemical Society. A.S. acknowledges the support from the French ANR (Project ProLiFic ANR-16-CE30-0009). E.V. thanks Roland Pellenq for pointing out Ref. \cite{Marder_13,bocquet_16}.

\bibliography{Biblio_Clogging.bib}

\end{document}